\newif\iftr
\newif\ifcnf

\trfalse
\cnftrue

\PassOptionsToPackage{prologue,svgnames}{xcolor}
\PassOptionsToPackage{numbers,sort&compress}{natbib}

\documentclass[conference,10pt,svgnames]{IEEEtran}

\usepackage[switch]{lineno}

\usepackage[hidelinks]{hyperref}
\usepackage{wrapfig}

\ifcnf
\fi

\makeatletter
\begingroup\endlinechar=-1\relax
       \everyeof{\noexpand}%
       \edef\x{\endgroup\def\noexpand\homepath{%
         \@@input|"kpsewhich --var-value=HOME" }}\x
\makeatother
\def\overleafhome{/tmp}

\usepackage{import}

\ifx\homepath\overleafhome

\newif\iftr     
\newif\ifconf   
\newif\ifnonb   

\usepackage[utf8]{inputenc}
\newcommand{\code}[1]{\texttt{#1}}
\usepackage[binary-units=true]{siunitx}
\usepackage{xspace}
\usepackage[svgnames]{xcolor}

\usepackage{collab}

\usepackage[numbers,sort&compress]{natbib}

\usepackage{soul}
\soulregister\cite7
\soulregister\autoref7
\soulregister\code7
\soulregister\toolname7
\soulregister\ref7
\soulregister\pageref7
\usepackage{hhline}
\definecolor{lightyellow}{RGB}{250, 250, 180}
\definecolor{HLYELLOW}{RGB}{240, 127, 0}
\definecolor{hlyellow}{RGB}{240, 127, 0}
\sethlcolor{lightyellow}

\usepackage{algorithm}
\usepackage{algpseudocode}
\usepackage{pifont}
\usepackage{amsmath}

\algdef{SE}[FOREACHIN]{ForEachIn}{EndForEachIn}[2]{\algorithmicfor\ \textbf{each}\ #1\ \textbf{in}\ #2\ \textbf{do}}{\algorithmicend\ \algorithmicfor}%
\algdef{SE}[FOREACHP]{ForEachP}{EndForEachP}[1]{\algorithmicfor\ \textbf{each}\ #1\ \textbf{in parallel do}}{\algorithmicend\ \algorithmicfor}%
\algdef{SE}[FOREACHINP]{ForEachInP}{EndForEachInP}[2]{\algorithmicfor\ \textbf{each}\ #1\ \textbf{in}\ #2\ \textbf{in parallel do}}{\algorithmicend\ \algorithmicfor}%
\algdef{SE}[FORMOD]{ForMod}{EndForMod}[3]{\algorithmicfor\ #1\ \textbf{from}\ #2\ \textbf{to}\ #3\ \textbf{do}}{\algorithmicend\ \algorithmicfor}%
\algdef{SE}[FORMOD]{ForModS}{EndForModS}[4]{\algorithmicfor\ #1\ \textbf{from}\ #2\ \textbf{to}\ #3\ \textbf{step}\ #4\ \textbf{do}}{\algorithmicend\ \algorithmicfor}%
\algdef{SE}[FORMODP]{ForModPS}{EndForModPS}[3]{\algorithmicfor\ #1\ \textbf{from}\ #2\ \textbf{to}\ #3\ \textbf{in parallel do}}{\algorithmicend\ \algorithmicfor}%
\algdef{SE}[FORMODP]{ForModP}{EndForModP}[4]{\algorithmicfor\ #1\ \textbf{from}\ #2\ \textbf{to}\ #3\ \textbf{step} #4\ \textbf{in parallel do}}{\algorithmicend\ \algorithmicfor}%

\algdef{SE}[VARIABLES]{Variables}{EndVariables}{\algorithmicvariables}
  {\algorithmicend\ \algorithmicvariables}
  \algnewcommand{\algorithmicvariables}{\textbf{global}}
\algnewcommand{\LineComment}[1]{\State \(\triangleright\) #1}
\algnewcommand{\And}{\textbf{and}\xspace}

\usepackage{tikz}
\usepackage{pifont}
\usetikzlibrary{shapes}

\DeclareRobustCommand*\circledColor[2]{\tikz[baseline=(char.base)]{
    \node[shape=circle,fill=#2,draw=#2,inner sep=1pt] (char) {\textcolor{white}{\small\textbf{#1}}};}}
\DeclareRobustCommand*\circledColorSmall[2]{\tikz[baseline=(char.base)]{
    \node[shape=circle,fill=#2,draw=#2,inner sep=0pt] (char) {\textcolor{white}{\footnotesize\textbf{#1}}};}}

\DeclareRobustCommand*\circledColorX[1]{\tikz[baseline=(char.base)]{
    \node[shape=circle,fill=#1,draw=#1,inner sep=2pt] {}
  }
}

\usepackage{listings}
\newfloat{codeblock}{H}{myc}
\definecolor{darkblue}{rgb}{0,0,.6}
\definecolor{darkred}{rgb}{.6,0,0}
\definecolor{darkgreen}{rgb}{0,.5,0}
\definecolor{red}{rgb}{.98,0,0}
\definecolor{gray}{rgb}{.6,.6,.6}
\definecolor{newgreen}{RGB}{169,209,142}
\definecolor{newpurple}{RGB}{237,134,254}
\definecolor{neworange}{RGB}{244,177,131}
\definecolor{newyellow}{RGB}{255,217,102}
\else

\newif\iftr     
\newif\ifconf   
\newif\ifnonb   

\usepackage[utf8]{inputenc}
\newcommand{\code}[1]{\texttt{#1}}
\usepackage[binary-units=true]{siunitx}
\usepackage{xspace}
\usepackage[svgnames]{xcolor}

\usepackage{collab}

\usepackage[numbers,sort&compress]{natbib}

\usepackage{soul}
\soulregister\cite7
\soulregister\autoref7
\soulregister\code7
\soulregister\toolname7
\soulregister\ref7
\soulregister\pageref7
\usepackage{hhline}
\definecolor{lightyellow}{RGB}{250, 250, 180}
\definecolor{HLYELLOW}{RGB}{240, 127, 0}
\definecolor{hlyellow}{RGB}{240, 127, 0}
\sethlcolor{lightyellow}

\usepackage{algorithm}
\usepackage{algpseudocode}
\usepackage{pifont}
\usepackage{amsmath}

\algdef{SE}[FOREACHIN]{ForEachIn}{EndForEachIn}[2]{\algorithmicfor\ \textbf{each}\ #1\ \textbf{in}\ #2\ \textbf{do}}{\algorithmicend\ \algorithmicfor}%
\algdef{SE}[FOREACHP]{ForEachP}{EndForEachP}[1]{\algorithmicfor\ \textbf{each}\ #1\ \textbf{in parallel do}}{\algorithmicend\ \algorithmicfor}%
\algdef{SE}[FOREACHINP]{ForEachInP}{EndForEachInP}[2]{\algorithmicfor\ \textbf{each}\ #1\ \textbf{in}\ #2\ \textbf{in parallel do}}{\algorithmicend\ \algorithmicfor}%
\algdef{SE}[FORMOD]{ForMod}{EndForMod}[3]{\algorithmicfor\ #1\ \textbf{from}\ #2\ \textbf{to}\ #3\ \textbf{do}}{\algorithmicend\ \algorithmicfor}%
\algdef{SE}[FORMOD]{ForModS}{EndForModS}[4]{\algorithmicfor\ #1\ \textbf{from}\ #2\ \textbf{to}\ #3\ \textbf{step}\ #4\ \textbf{do}}{\algorithmicend\ \algorithmicfor}%
\algdef{SE}[FORMODP]{ForModPS}{EndForModPS}[3]{\algorithmicfor\ #1\ \textbf{from}\ #2\ \textbf{to}\ #3\ \textbf{in parallel do}}{\algorithmicend\ \algorithmicfor}%
\algdef{SE}[FORMODP]{ForModP}{EndForModP}[4]{\algorithmicfor\ #1\ \textbf{from}\ #2\ \textbf{to}\ #3\ \textbf{step} #4\ \textbf{in parallel do}}{\algorithmicend\ \algorithmicfor}%

\algdef{SE}[VARIABLES]{Variables}{EndVariables}{\algorithmicvariables}
  {\algorithmicend\ \algorithmicvariables}
  \algnewcommand{\algorithmicvariables}{\textbf{global}}
\algnewcommand{\LineComment}[1]{\State \(\triangleright\) #1}
\algnewcommand{\And}{\textbf{and}\xspace}

\usepackage{tikz}
\usepackage{pifont}
\usetikzlibrary{shapes}

\DeclareRobustCommand*\circledColor[2]{\tikz[baseline=(char.base)]{
    \node[shape=circle,fill=#2,draw=#2,inner sep=1pt] (char) {\textcolor{white}{\small\textbf{#1}}};}}
\DeclareRobustCommand*\circledColorSmall[2]{\tikz[baseline=(char.base)]{
    \node[shape=circle,fill=#2,draw=#2,inner sep=0pt] (char) {\textcolor{white}{\footnotesize\textbf{#1}}};}}

\DeclareRobustCommand*\circledColorX[1]{\tikz[baseline=(char.base)]{
    \node[shape=circle,fill=#1,draw=#1,inner sep=2pt] {}
  }
}

\definecolor{darkblue}{rgb}{0,0,.6}
\definecolor{darkred}{rgb}{.6,0,0}
\definecolor{darkgreen}{rgb}{0,.5,0}
\definecolor{red}{rgb}{.98,0,0}
\definecolor{gray}{rgb}{.6,.6,.6}
\definecolor{newgreen}{RGB}{169,209,142}
\definecolor{newpurple}{RGB}{237,134,254}
\definecolor{neworange}{RGB}{244,177,131}
\definecolor{newyellow}{RGB}{255,217,102}

\fi
\collabAuthor{copik}{blue}{Marcin}
\collabAuthor{lex}{violet}{lex}
\collabAuthor{htor}{magenta}{Torsten}
\collabAuthor{todos}{red}{TODO}
\collabAuthor{konst}{green}{Konstantin}

\ifx\homepath\overleafhome
  \usepackage[frozencache=true,cachedir=.,outputdir=.]{minted}
\else
  \usepackage[frozencache=true,cachedir=.,outputdir=.]{minted}
\fi

\usepackage{enumitem}

\usepackage{enumitem}
\usepackage{fontawesome}
\usepackage{adjustbox}
\usepackage{booktabs}
\usepackage{subfig}

\usepackage{moresize}

\usepackage{minted}

\newcommand{\toolname}{\emph{rFaaS}\xspace}

\usepackage[many]{tcolorbox}
\tcbuselibrary{skins}
\usetikzlibrary{calc}

\definecolor{myblue}{RGB}{0,163,243}
\definecolor{textnewGreen}{HTML}{66AE3E}
\definecolor{textNewViolet}{HTML}{905E96}
\definecolor{bgnewGreen}{HTML}{EBF4DE}
\newtcbox{\titleboxblue}{
  enhanced,
  colupper=white,
  colback=textNewViolet,
  fontupper=\bfseries\sffamily\small,
  size=small,
  baseline=4pt,
  nobeforeafter,
	left=0pt,
	right=0pt,
	top=-1pt,
	bottom=-1pt,
  frame code={
    \path[fill=textNewViolet] (frame.north west)
    -- ([xshift=2mm]frame.north east)
    -- (frame.south east)
    -- (frame.south west)
    -- (frame.north west)
      [sharp corners]-- cycle;
  }
}

\newtcbox{\titleboxgreen}{
  enhanced,
  colupper=white,
 colback=textnewGreen,
  fontupper=\bfseries\sffamily\large,
  size=small,
  baseline=4pt,
  nobeforeafter,
	left=0pt,
	right=0pt,
	top=0pt,
	bottom=-1pt,
  frame code={
    \path[fill=textnewGreen] (frame.north west)
    -- ([xshift=2mm]frame.north east)
    -- (frame.south east)
    -- (frame.south west)
    -- (frame.north west)
      [sharp corners]-- cycle;
  }
}

\makeatletter

\newtcolorbox{summaryblue}[1]{
  enhanced,
  skin=bicolor,
  arc=0pt,
	left=0pt,
	right=0pt,
	top=0pt,
	bottom=0pt,
  coltitle=white,
  colframe=textNewViolet,
  colback=textNewViolet!20,
  colbacklower=white,
  detach title,
  title={#1},
  before upper*={%
    \vskip-\dimexpr\kvtcb@boxsep+\kvtcb@top+.1pt
    \hspace*{-\dimexpr\kvtcb@boxsep+\kvtcb@leftupper+.1pt}%
    \expandafter\titleboxblue\expandafter{\tcbtitletext}
  }
}
\newtcolorbox{summarygreen}[1]{
  enhanced,
  skin=bicolor,
  arc=0pt,
	left=0pt,
	right=0pt,
	top=0pt,
	bottom=0pt,
  coltitle=white,
  colframe=textnewGreen,
  colback=bgnewGreen,
  colbacklower=white,
  detach title,
  title={#1},
  before upper*={%
    \vskip-\dimexpr\kvtcb@boxsep+\kvtcb@top+.1pt
    \hspace*{-\dimexpr\kvtcb@boxsep+\kvtcb@leftupper+.1pt}%
    \expandafter\titleboxgreen\expandafter{\tcbtitletext}
  }
}
\makeatother

\begin{document}

\title{rFaaS: Enabling High Performance Serverless with RDMA and Leases}

\author{
    \IEEEauthorblockN{
        Marcin Copik\IEEEauthorrefmark{1},
        Konstantin Taranov\IEEEauthorrefmark{2},
        Alexandru Calotoiu\IEEEauthorrefmark{1},
        Torsten Hoefler\IEEEauthorrefmark{1}
    }
    \IEEEauthorblockA{\IEEEauthorrefmark{1}Department of Computer Science,
      ETH Z{\"u}rich, Z{\"u}rich, Switzerland}
    \IEEEauthorblockA{\IEEEauthorrefmark{2}Microsoft}
    \IEEEauthorrefmark{1}firstname.lastname@inf.ethz.ch, \IEEEauthorrefmark{2}kotaranov@microsoft.com
}
\maketitle

\begin{abstract}
  High performance is needed in many computing systems, from batch-managed supercomputers to general-purpose cloud platforms.
  However, scientific clusters lack elastic parallelism, while clouds cannot offer competitive
  costs for high-performance applications.
  In this work, we investigate how modern cloud programming paradigms can bring the elasticity needed to allocate idle resources,
  decreasing computation costs and improving overall data center efficiency.
  Function-as-a-Service (FaaS) brings the pay-as-you-go execution of stateless functions, but its
  performance characteristics cannot match coarse-grained cloud and cluster allocations.
  %
  %
  To make serverless computing viable for high-performance and latency-sensitive applications, we present
  rFaaS, an RDMA-accelerated FaaS platform.
  We identify critical limitations of serverless
  - centralized scheduling and inefficient network transport -
  and improve the FaaS architecture with allocation leases
  and microsecond invocations.
   We show that our remote functions add only negligible overhead on top of the fastest available
  networks, and we decrease the execution latency by orders of magnitude compared to
  contemporary FaaS systems.
  Furthermore, we demonstrate the performance of rFaaS by evaluating real-world FaaS
  benchmarks and parallel applications.
  Overall, our results show that new allocation policies and remote memory access help FaaS
  applications achieve high performance and bring serverless computing to HPC.
\end{abstract}

\maketitle

\noindent \textbf{rFaaS Implementation}: \url{https://github.com/spcl/rFaaS}

\noindent \textbf{rFaaS Artifact:} \url{https://zenodo.org/record/7657524}

\section{Introduction}


\iftr
The high-performance computing landscape is dominated by the Message-Passing Interface (MPI),
a \emph{de facto} standard distributed programming paradigm.
With job batch scheduling~\cite{10.1007/3-540-63574-2_14}
and multithreaded frameworks for shared-memory programming,
MPI is the leading use case for clusters and supercomputers~\cite{10.1145/2063348.2063374}.
In the rigid HPC world, applications that exhibit varying parallelism achieve lower efficiency
because adapting resource allocation to changing requirements is heavily constrained~\cite{10.1007/BFb0022284,6008941}.
%
%
%
On the other hand, the cloud has brought a major innovation in the form of elastic resource management~\cite{Armbrust09abovethe}.
While cloud computing has the hardware capability to support high-performance workloads, it lacks
programming models needed for flexible parallelism.
Without such models, high-performance applications overprovision computing resources and increase the data center underutilization,
a problem that has a significant impact as
\emph{"increasing utilization by a few percentage points can save millions of dollars"}~\cite{10.1145/2741948.2741964}.
Furthermore, high-performance computing units in the cloud can be substantially more expensive
than existing supercomputing resources\footnote{For example, nodes of the Piz Daint supercomputer~\cite{pizDaint}
start at 0.53 CHF (approximately 0.55 USD) per nodehour~\cite{pizDaintPricing}.
Comparable pay-as-you-go cloud instances cost at least 1.5 and as high as 4.08 USD~\cite{awsEC2Pricing,azureVMPricing}.
While cloud instances can be cheaper, these plans require longer-term allocations for many months and provide no flexibility.}.
%
While HPC systems offer cheap resources, their rigid structure limits efficiency.
Conversely, high-performance applications cannot use elastic cloud resources due to the lack of an efficient framework.
\else
The high-performance computing landscape is dominated by the Message-Passing Interface (MPI),
the \emph{de facto} standard distributed programming paradigm.
With job batch scheduling
and shared--memory frameworks for multithreading,
MPI is the leading use case for clusters and supercomputers~\cite{10.1145/2063348.2063374}.
In the rigid HPC world, applications with varying parallelism achieve lower efficiency
because adapting resource allocation to changing requirements is heavily constrained~\cite{10.1007/BFb0022284,6008941}.
%
%
%
On the other hand, the cloud has brought a major innovation in elastic resource management.
While cloud computing has the hardware capability to support high-performance workloads, it lacks
programming models for flexible parallelism.
Thus, high-performance applications overprovision computing resources and increase the data center underutilization,
a problem that has a significant impact as
\emph{"increasing utilization by a few percentage points can save millions of dollars"}~\cite{10.1145/2741948.2741964}.
Furthermore, HPC units in the cloud can be substantially more expensive
than existing supercomputing resources\footnote{For example, nodes of the Piz Daint supercomputer
start at 0.53 CHF (approximately 0.55 USD) per nodehour.
Comparable pay-as-you-go cloud instances cost at least 1.5 and as high as 4.08 USD.
Cheaper cloud instances require longer-term allocations for many months and provide no flexibility.}.
%
The rigid structure of supercomputing systems limits their efficiency, and HPC applications cannot
benefit from cloud elasticity due to lack of flexible frameworks.
\fi

\begin{figure}[tb!]
	\centering
  \includegraphics[width=1.0\linewidth]{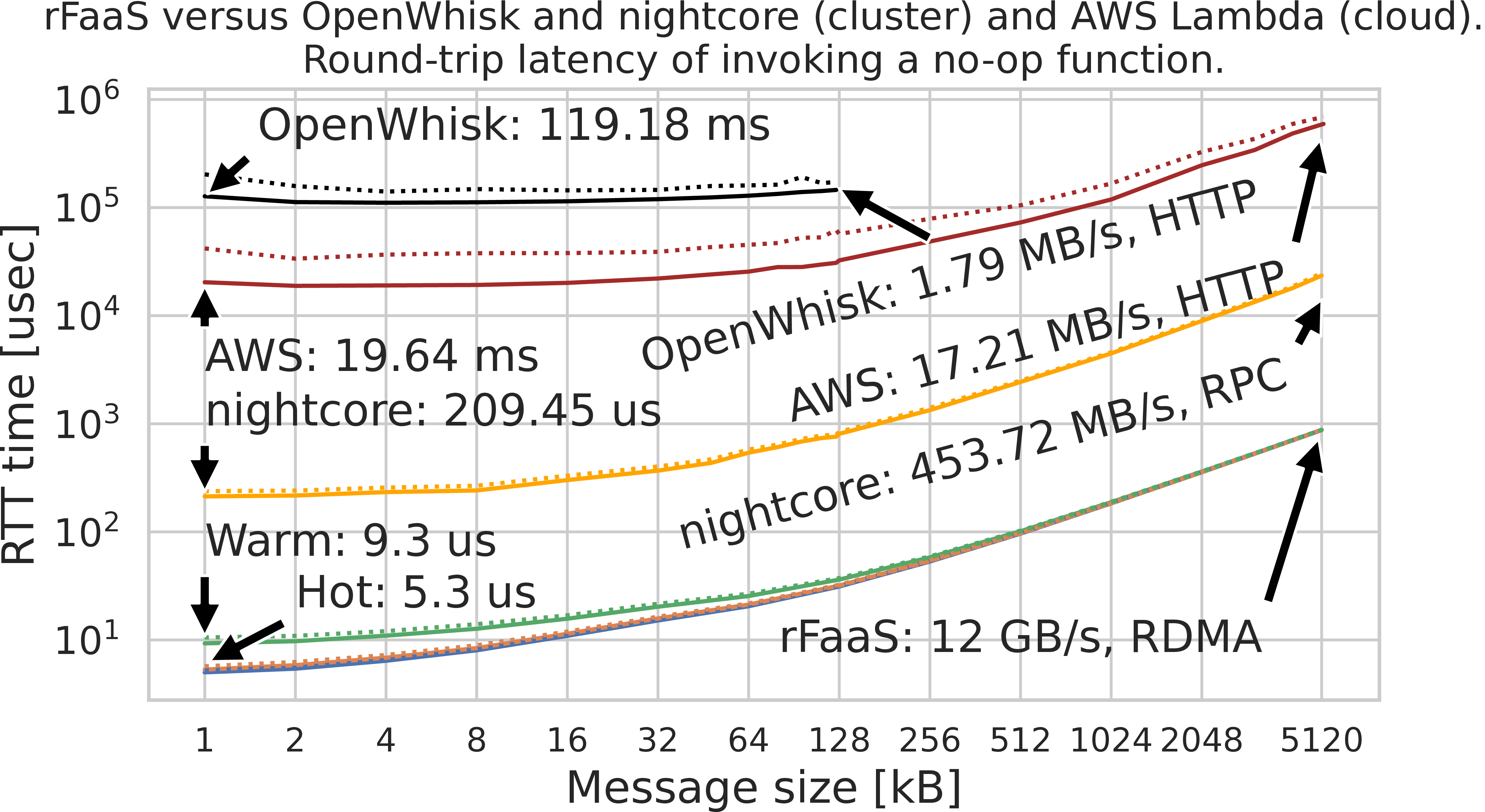}
  \vspace{-1em}
  \caption{
    \textbf{The remote invocations of an empty C++ function on
    serverless platforms and \toolname: median (solid), and 99th latency (dashed) for a single function
  (details in Sec.~\ref{sec:evaluation_payload}).}
  }
  \vspace{-2.5em}
  \label{fig:evaluation_faas}
\end{figure}

\iftr
The performance gap between public clouds and supercomputing data centers has been shrinking over time.
The performance and attractiveness of \emph{Infrastructure-as-a-Service (IaaS)} resources
has improved~\cite{6114491,10.1145/3150224,7573827,10.1145/3140607.3050765}.
Furthermore, virtual machines and containers, the important virtualization solutions
in cloud systems, have been found to be an efficient abstraction level for high-performance applications~\cite{10.1145/1183401.1183421,10.1007/978-3-319-27308-2_65}.   
The cloud's elastic scalability fueled new applications in massive data analytics,
image and video processing, and large-scale machine learning.
Thus, parallel applications running in private data centers and scientific supercomputers started taking advantage of vast cloud resources~\cite{5754008,10.1145/3150224,10.1145/2063384.2063462}.
%
%
However, high-performance applications must provision computational resources for peak demand,
satisfying strict performance and \emph{Quality-of-Service (QoS)} requirements.
Thus, they employ complex and domain-specific optimizations such as inference-focused systems~\cite{234998,9355312}
or resort to overprovisioning and underutilizing computing resources.
As a result, the utilization of public and private clouds, such as supercomputers,
has always been low for both computing and memory resources (Sec.~\ref{sec:background_resources}).
HPC users want to achieve higher efficiency with the elasticity of the cloud without sacrificing performance guarantees.
Unfortunately, the question of incorporating elastic resources into frameworks such as MPI remains open due to a lack of an appropriate programming model.
\else
The performance gap between public clouds and HPC data centers has been shrinking over time.
The performance and attractiveness of \emph{Infrastructure-as-a-Service (IaaS)} resources
has improved~\cite{6114491,10.1145/3150224,7573827}.
Furthermore, virtual machines and containers have been found to be an efficient abstraction
level for high-performance applications~\cite{10.1145/1183401.1183421,10.1007/978-3-319-27308-2_65}.   
Thus, parallel applications running in private data centers and scientific supercomputers started
taking advantage of vast cloud resources~\cite{10.1145/3150224}.
%
%
However, high--performance applications must provision resources for the peak demand.
Thus, they employ complex and domain-specific optimizations or resort to overprovisioning and underutilizing computing resources.
As a result, the resource utilization in public clouds and private data centers, such as supercomputers,
has always been low for both computing and memory resources (Sec.~\ref{sec:background_resources}).
HPC users want to use flexible allocations and achieve higher efficiency without sacrificing performance guarantees.
Unfortunately, the question of incorporating elastic resources into HPC frameworks
remains open.

\fi



%
%
\iftr
To bridge the gap between supercomputers and the cloud, we must incorporate elastic allocations into supercomputers and decrease HPC computation costs in the cloud.
This new programming model would need to satisfy two major requirements.
First, high-performance computing units should be immediately available to address workload and load balance changes. 
However, acquiring new \emph{Infrastructure-as-a-Service (IaaS)} resources in the cloud is time-consuming.
Even as of 2021, the time needed to allocate a virtual machine is counted in minutes,
not seconds~\cite{hao2021empirical}.
Similarly, HPC batch systems do not support dynamic resource reallocation.
Second, the model should allow users to allocate and use idle resources effectively.
These can be offered at a significantly lower cost since decreasing resource waste increases the overall efficiency of a data center.
However, the rapid and frequent utilization changes on many such platforms (Fig.~\ref{fig:daint_stats})
indicate that persistent and long-running allocations cannot address these idleness gaps.
The programming model must support fine-grained and ephemeral allocations to take advantage of
resources available only for a short time.
\else
To bridge the gap between HPC clusters and the cloud, we must incorporate elastic allocations
into supercomputers and decrease HPC computation costs in the cloud.
This new programming model would need to satisfy two essential requirements.
First, high-performance computing units should be immediately available to address workload and
load balance changes. 
However, even as of 2021, acquiring new \emph{Infrastructure-as-a-Service (IaaS)} resources in the
cloud takes minutes, not seconds~\cite{hao2021empirical}.
Similarly, HPC batch systems do not support dynamic resource reallocation.
Second, the model should allow users to effectively use idle resources.
These can be offered at a significantly lower cost since decreasing resource waste increases
the overall efficiency of a data center.
However, the rapid and frequent utilization changes on many such platforms (Fig.~\ref{fig:daint_stats})
indicate that persistent and long-running allocations cannot address these idleness gaps.
Thus, the programming model should use fine-grained and ephemeral workers to allocate
resources with short availability.
\fi

%
\emph{Function-as-a-Service (FaaS)} is a new cloud paradigm combining the full elasticity
of cloud resources with a maximally simplified programming model: users program
stateless functions and the cloud completely manages their scheduling.
Thanks to the fine-grained parallelism and the pay-as-you-go billing system,
serverless functions could become a solution for all tasks that
benefit from an elastic allocation of computing resources.
\iftr
Functions are used as costly but flexible elastic workers to fulfill
\emph{Service Level Objective (SLO)} requirements~\cite{8814535,234998},
and they could become the implementation of \emph{HPC-as-a-Service}~\cite{6280554}.
However, functions are not yet ready for 
high-performance applications.
%
For FaaS to become a viable programming model for HPC,
it must overcome crucial performance challenges (Table~\ref{tab:faas_limitations}).
\else
Functions are used as elastic workers to fulfill
\emph{Service Level Objective (SLO)} requirements~\cite{8814535},
and they could implement the \emph{HPC-as-a-Service}~\cite{6280554}.
However, functions must overcome crucial performance challenges
to become a viable programming model for HPC (Table~\ref{tab:faas_limitations}).
\fi


We address these challenges in \textbf{\toolname{}}, an RDMA-capable serverless platform
tailored to the requirements of HPC applications (Sec.~\ref{sec:design}).
We define new \textbf{RDMA abstractions} that hide the network stack complexity and
preserve the elasticity and isolation of serverless.
We improve serverless architecture with three innovations that integrate into existing designs.
First, \toolname{} employs a new resource management policy where \textbf{leases} replace centralized
placement of invocations.
Instead of routing every function to the same warm containers, leases allow to skip the control logic.
Then, we accelerate the serverless system by \textbf{reducing its invocation path} for high--priority and
low--latency tasks: \toolname{} invocations are handled directly between the client and a function executor.
Finally, to achieve microsecond latency invocations, we replace HTTP and REST interfaces with an \textbf{RDMA function dispatch protocol}
that removes the milliseconds of OS latency~\cite{guo2016rdma}.
\iftr
We help to integrate FaaS executions into high-performance
applications
and show \textbf{hot} invocations with an overhead
of a little over 300 nanoseconds on top of the fastest network (Fig.~\ref{fig:evaluation_faas}).
\else
We show \textbf{hot} invocations with an overhead
of a little over 300 nanoseconds on top of the fastest network (Fig.~\ref{fig:evaluation_faas}).
\fi

\iftr
To incorporate serverless computing into high-performance applications,
we present \textbf{a C++ programming model} for straightforward integration of \toolname{}
functions into new and existing C++ codebases (Sec.~\ref{sec:integration}).
%
Our work is a major step towards increasing efficiency by using idle and ephemeral
resources for tasks demanding high performance.
We demonstrate the elasticity, efficiency, and performance of \toolname{} with an evaluation
of microbenchmarks and real-world serverless functions (Sec.~\ref{sec:evaluation}).
\else
We present a \textbf{C++ programming model} for straightforward integration of \toolname{}
functions into high-performance applications (Sec.~\ref{sec:integration}).
%
Our work is a major step towards increasing efficiency by using idle and ephemeral
resources for tasks demanding high performance.
We demonstrate the elasticity, efficiency, and performance of \toolname{} with an evaluation
of microbenchmarks, functions, and HPC applications (Sec.~\ref{sec:evaluation}).
\fi

\ifcnf
\setlength{\skip\footins}{8pt}
\fi

Our paper makes the following contributions:
\begin{itemize}[noitemsep,nosep]
  \item We present the design and open-source implementation
  of the first RDMA-capable serverless platform, including
(1) new FaaS resource management and
(2) a novel, low-latency, and zero-copy \emph{hot} type of serverless invocations.
  \item We conduct an experimental verification against state-of-the-art open-source and
    commercial serverless platforms summarized in Fig.~\ref{fig:evaluation_faas} and show that \toolname{}
has a median overhead over pure RDMA transmission of little over 300 ns
and achieves the available link bandwidth.
  \item We demonstrate \toolname{} usability with real-world serverless functions
    and show how the invocation latency is sufficient to accelerate HPC applications.
\end{itemize}

\begin{table}
  \footnotesize
  \begin{adjustbox}{max width=\linewidth}
  \begin{tabular}{lcc}
		\toprule
    \textbf{Requirements} & \textbf{rFaaS} & \textbf{Other solutions.}\\
		\midrule
    Low-latency invocations   & \faThumbsOUp    & Nightcore~\cite{nightcore}                  \\
    Direct allocations        & \faThumbsOUp    & \faThumbsDown                               \\
    High-speed networks       & \faThumbsOUp    & \faThumbsDown                               \\
    Decentralized scheduling  & \faThumbsOUp    & Wukong~\cite{wukong}, Archipelago~\cite{archipelago}              \\
    Efficient workflows       & \faThumbsUp     & SAND~\cite{10.5555/3277355.3277444}, Wukong~\cite{wukong}, Cloudburst~\cite{10.14778/3407790.3407836}.                    \\
    Direct communication      & \faThumbsUp     & Boxer~\cite{wawrzoniak2021boxer}            \\
		\midrule
    Fast and shared storage   & \multicolumn{2}{c}{Open problem.}                             \\
    Affordable costs          & \multicolumn{2}{c}{Open problem.}                             \\
    Consistent performance    & \multicolumn{2}{c}{Open problem.}                             \\
		\bottomrule
  \end{tabular} 
  \end{adjustbox}
  \caption{ 
    \textbf{\toolname{} solves (\faThumbsOUp) and enables solutions (\faThumbsUp)
    to the major challenges of high-performance FaaS~\cite{DBLP:journals/corr/abs-1902-03383,copik2021sebs,jiang2021towards,lopez2021serverless}.}
  }
  \vspace{-2em}
	\label{tab:faas_limitations}
\end{table}

\section{Background}

\iftr
\begin{figure*}[tbh!]
	\centering
    \subfloat[Idle CPU cores rate (\%).
    ]{%
      \includegraphics[width=0.45\linewidth]{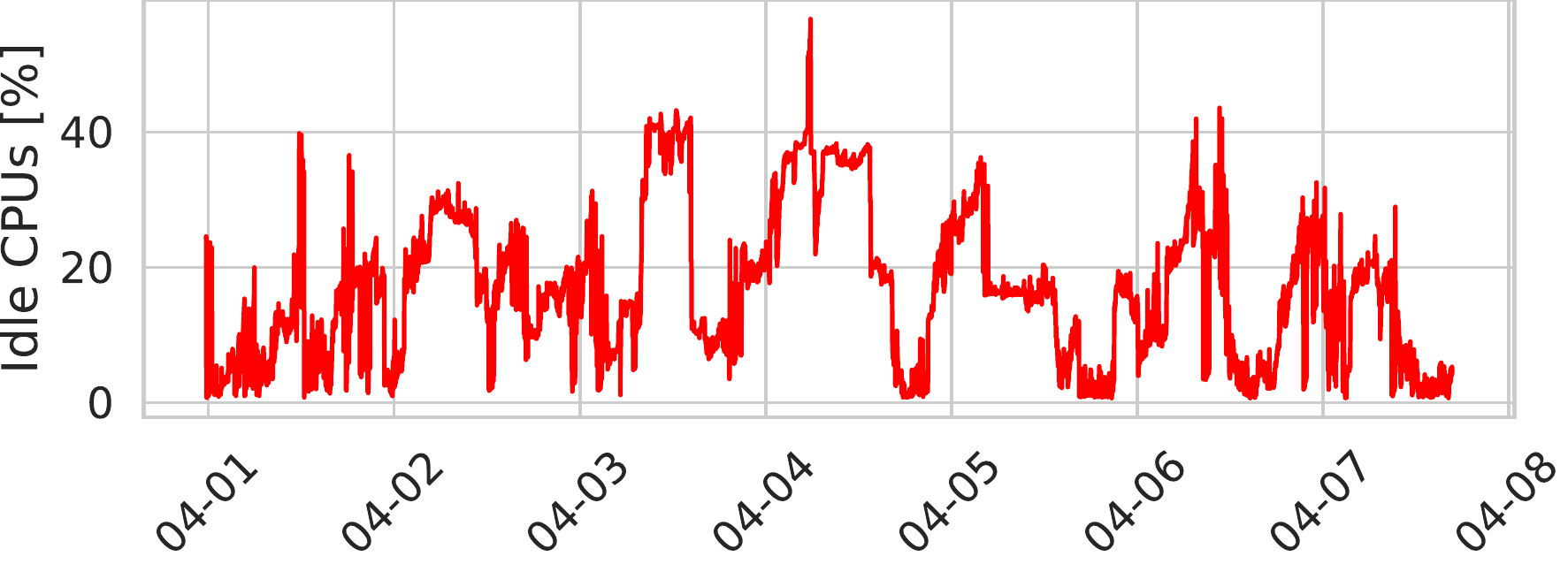}\label{fig:daint_cpu}
    }
	\hfill
		\subfloat[Free memory rate (\%).
		]{%
      \includegraphics[width=0.45\linewidth]{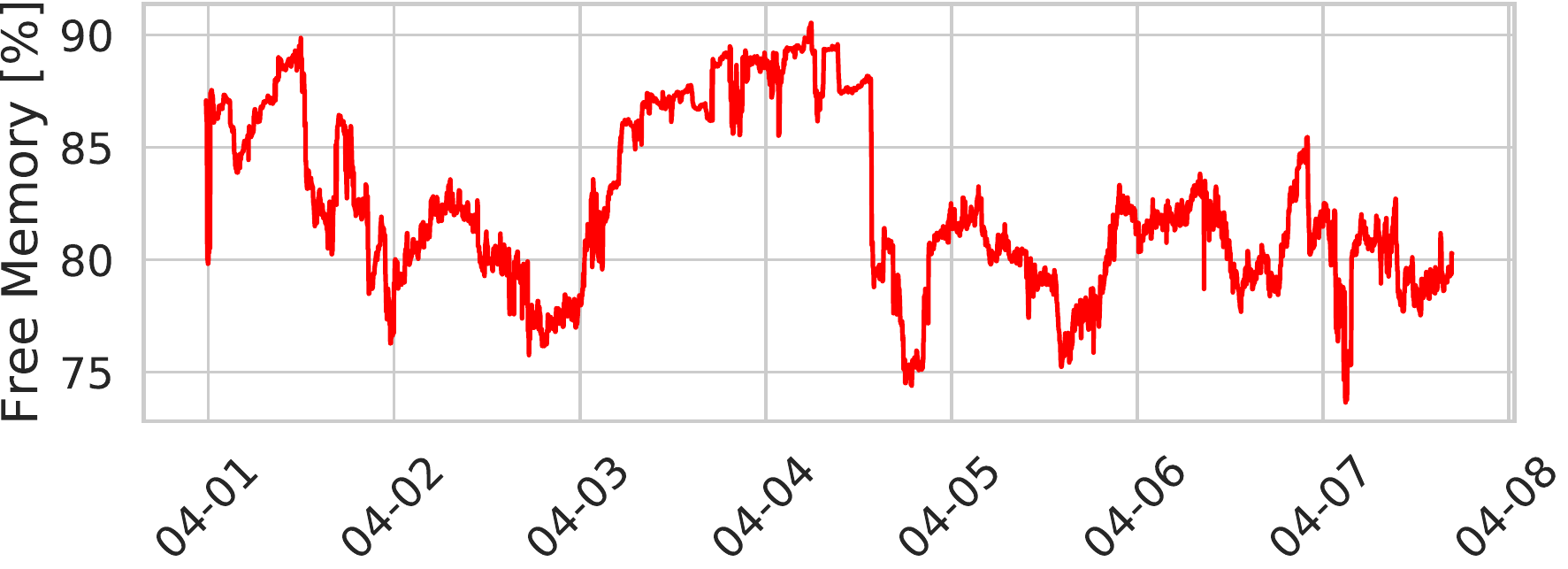}
      \label{fig:daint_memory}
		}
  \vspace{-0.5em}
  \caption{Utilization of the Piz Daint supercomputer for one week (31.03-7.04 in 2021): querying the batch manager SLURM with a one-minute interval. Many nodes are idle for a short time, and this HPC cluster has large pools of unused memory.}
  \vspace{-1.5em}
  \label{fig:daint_stats}
\end{figure*}
\fi

\label{sec:background}
\toolname{} solves the utilization problems of data centers by 
identifying the opportunity to reuse idle resources (Sec.~\ref{sec:background_resources}).
At the same time, modern FaaS platforms are too constrained (Sec.~\ref{sec:background_faas})
to take advantage of high-speed networks and remote memory operations (Sec.~\ref{sec:background_rdma}),
motivating improvements to the serverless architecture (Sec.~\ref{sec:design}).

\ifcnf
\begin{figure}[tbh!]
	\centering
    \subfloat[Idle CPU cores rate (\%).
    ]{%
      \includegraphics[width=0.49\linewidth]{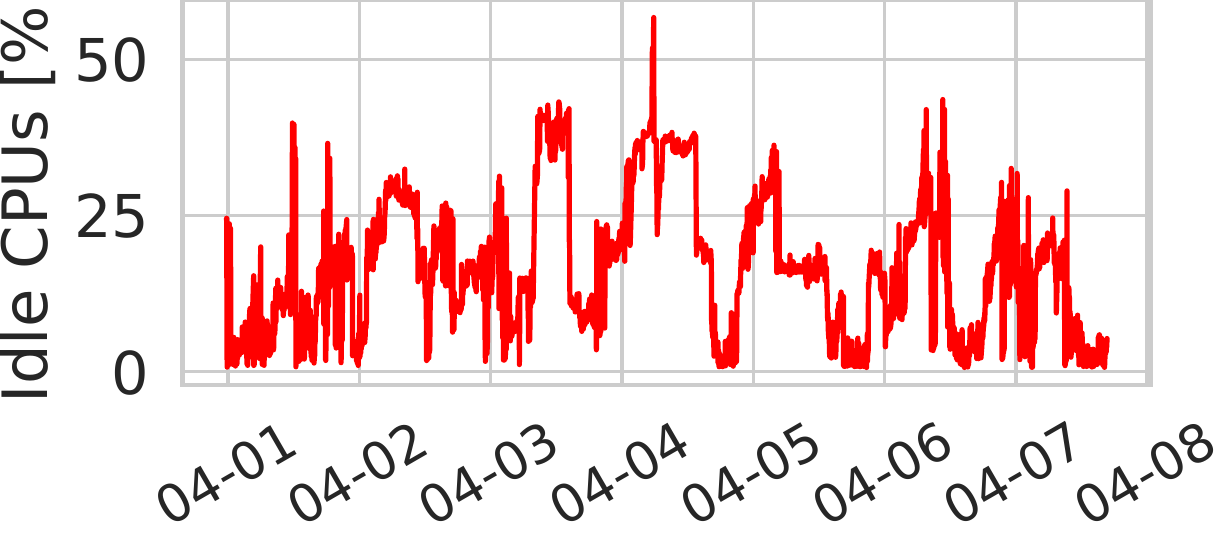}
      \label{fig:daint_cpu}
    }
	\hfill
		\subfloat[Free memory rate (\%).
		]{%
      \includegraphics[width=0.48\linewidth]{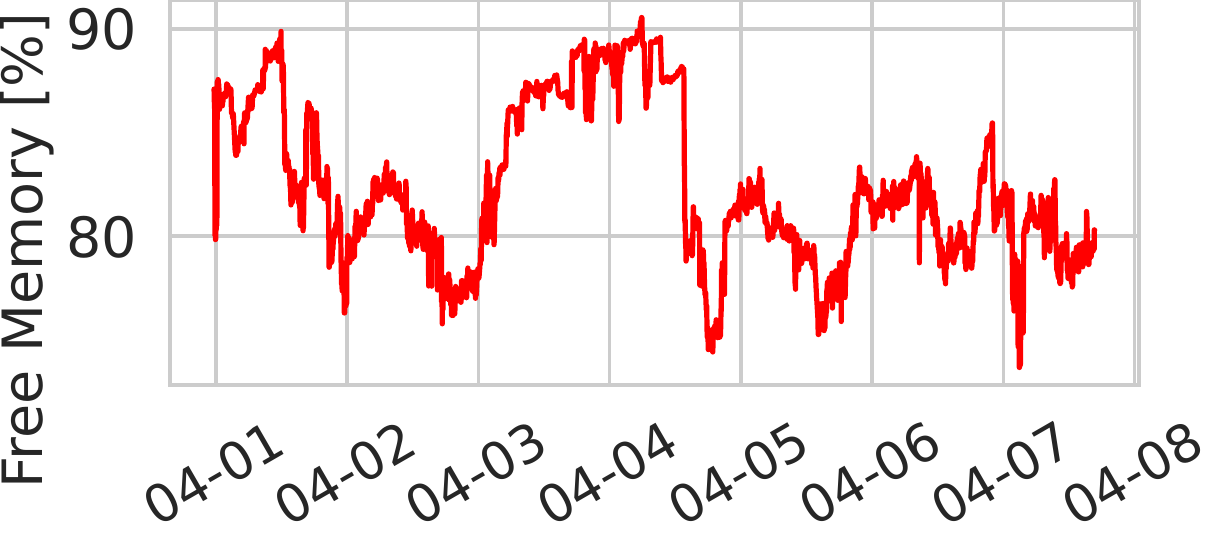}
      \label{fig:daint_memory}
		}
  \caption{\textbf{Piz Daint supercomputer utilization on 31.03-7.04 2021: querying SLURM with a one-minute interval.}}
  \vspace{-1.5em}
  \label{fig:daint_stats}
\end{figure}
\fi

\subsection{Resource Utilization}
\label{sec:background_resources}

\iftr
At the advent of cloud computing, server utilization
has been estimated to
be just 5-20\% (2008, 2010)~\cite{10.1145/1721654.1721672, kaplan2008revolutionizing}.
%
The latter study indicated that 30\% of servers
in data centers
do not perform any work.
%
Low utilization negatively affects capital investments through wasted 
resources,
and increases operating costs,
as the energy usage of servers doing little and no work is more than
50\% of their peak power consumption~\cite{4404806}.

\paragraph{Resources}
Datacenter resources are heavily underutilized.
In data centers, the CPU utilization was
5-10\% (Yahoo!, 2008-2009)~\cite{5493490} and 18\% (IBM, 2009-2011)~\cite{6253523}.
Similarly, the utilization of \emph{private clouds} was 10\% (2013)~\cite{7152512}
and less than 20\% for 80\% of clusters (2016)~\cite{10.1145/2987550.2987584}.
In Google data centers, the CPU utilization does not exceed 60\% (2011)~\cite{10.1145/2391229.2391236,6529276},
and while recent results indicate an increase through the "best-effort" batch jobs,
the utilization still does not exceed 60\% (2019)~\cite{10.1145/3342195.3387517}.
In the cloud, the average CPU utilization of virtual machines was
4-17\% (AWS, 2011)~\cite{6118751},
and a median for half of the instances was below 10\%
(Azure, 2016-2017)~\cite{10.1145/3132747.3132772}.
In the Alibaba cloud, the utilization per machine was within 40\% (2017)~\cite{8258257}
and consistently low for a majority of containers~\cite{10.1145/3267809.3267830}.
In highly \mbox{competitive} and batch-managed supercomputers, the average node utilization
varies between 80\% and 94\%~\cite{10.1007/978-3-642-35867-8_14,10.5555/3433701.3433812,DBLP:journals/corr/JonesWIDSPGFSBK17}.
We observed similar node utilization in the supercomputing system Piz Daint~\cite{pizDaint}
(Fig.~\ref{fig:daint_cpu}).
Futhermore, on average three-quarters of memory in an HPC node is not utilized~\cite{10.1145/3352460.3358267}.
These large quantities of idle memory (Fig.~\ref{fig:daint_memory}) offer possibility to host
warm state of ephemeral workers.
Many nodes are idle for a short time, and this HPC cluster has large pools of unused memory.

\noindent \textbf{Observation \#1} CPU and memory resource usage might be low even on allocated
servers. Therefore, users would benefit from \emph{reselling} unused cycles and the
accompanying infrastructure to host serverless invocations.

\paragraph{Workload variability}
Studies analyzing cloud and cluster workloads reveal significant temporal, spatial, and diurnal
variability.
In the Alibaba cloud data center~\cite{8258257}, some instances use less than 10\% and others
60-90\% of their CPU,
and multi-core containers underutilize CPUs because of the temporal workload
variability~\cite{10.1145/3267809.3267830}.
Analysis of enterprise workloads reveals heavy skew in the workloads,
where CPU usage is unpredictable and dynamic for 20\% of VMs~\cite{7152512},
and peak usage is much higher than 90th and 99th percentile~\cite{267668},
forcing the system to be ready to release significant resources for the peak traffic.
While overprovisioned resources could be reclaimed (Sec.~\ref{sec:related_work}),
it usually requires extensive profiling and application requirements classification.
Furthermore, the prime causes of overprovisioning are the low-latency demands in application SLOs,
which force reclaimed resources to be transient and easily retrievable by the application.
In the supercomputing system Piz Daint, the idle resources are available only for a short time,
requiring reclaiming approaches to support only short-running workloads (Fig.~\ref{fig:daint_cpu}).

\noindent \textbf{Observation \#2:}
Variable workloads make reallocations and
under-provisioning challenging. Stateless and short-lived functions are a natural fit for the idea of
\emph{opportunistic} computing~\cite{10.1145/3477132.3483580}, and \toolname{} can employ \emph{ephemeral} cloud resources.

\else
Low resource utilization has always affected data centers and it had a vast impact on the
financial efficiency of the system: wasted capital investments into idle resources
and increased operating costs, as the energy usage of servers doing little and no work is more than
50\% of their peak power consumption~\cite{4404806}.
In highly \mbox{competitive} and batch-managed supercomputers, the average utilization of nodes varies between 80\% and 94\%~\cite{10.1007/978-3-642-35867-8_14,10.5555/3433701.3433812,DBLP:journals/corr/JonesWIDSPGFSBK17}.
Furthermore, on average three-quarters of the memory in HPC nodes is not utilized~\cite{10.1145/3352460.3358267}.
We observed similar underutilization problems in the supercomputing system Piz Daint.
Since the idle nodes are available for a short time (Fig.~\ref{fig:daint_cpu}),
opportunistic reuse for other computations must be constrained to short-running workloads.
To support incoming large-scale jobs, reclaimed resources must be transient and easily retrievable
by the batch system.
Fortunately, large quantities of idle node memory open the possibility of hosting the warm state of
ephemeral workers (Fig.~\ref{fig:daint_memory}).
\vspace{-0.5em}
\begin{summaryblue}{Observation}
\noindent
Stateless and short-lived functions are a natural fit for
\emph{opportunistic} computing,
and \toolname{} can employ \emph{ephemeral} HPC resources.
\end{summaryblue}
\fi

\begin{figure}[t]
	\centering
  \includegraphics[width=1.0\linewidth]{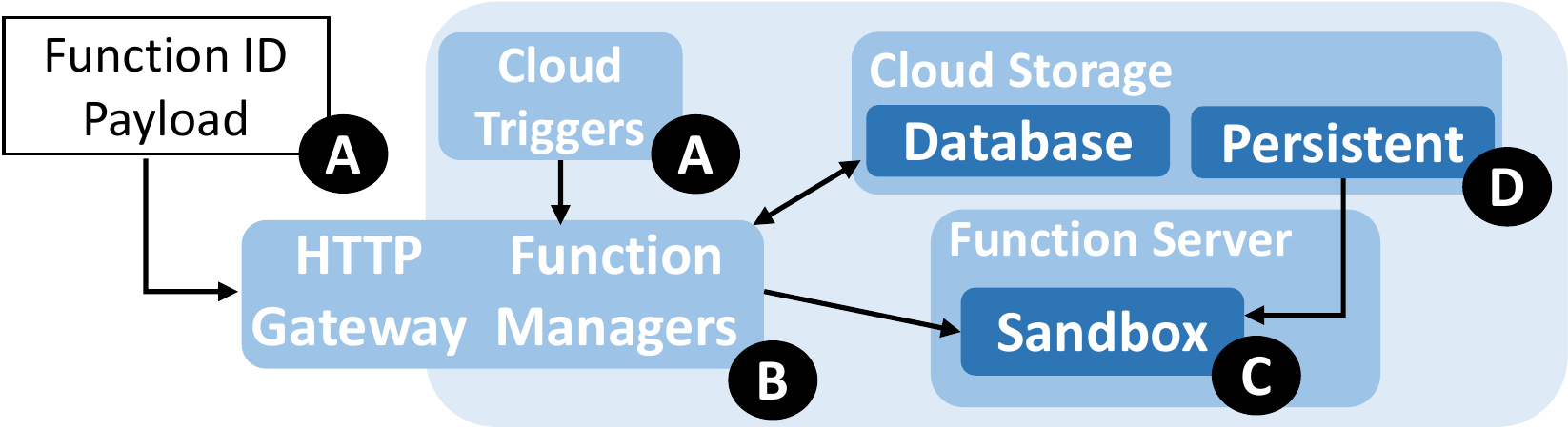}
  \caption{\textbf{A high-level view of the FaaS architecture.}}
  \vspace{-1.5em}
  \label{fig:faas_model}
\end{figure}

\ifcnf
\vspace{-1em}
\fi
\subsection{FaaS Computing}
\label{sec:background_faas}
Function-as-a-Service (FaaS) is a cloud service concerned with executing stateless and
short-running functions.
The serverless functions are dynamically allocated in the cloud,
and the users are freed from the usual responsibilities of managing resources.
The cloud provider charges users only for the time and resources used in a function
execution, and applications with irregular or infrequent workloads can benefit from the elastic
allocation of computing resources and the pay-as-you-go billing system.
For a cloud operator, the fine-grained executions provide an opportunity to increase
system efficiency through oversubscription and flexible scheduling.

\iftr
\paragraph{Platform}
We characterize the FaaS platforms with a high-level overview presented in Fig.~\ref{fig:faas_model}
and refer interested readers to a wider
discussion in the literature~\cite{DBLP:journals/corr/abs-1902-03383,copik2021sebs,10.5555/3277355.3277369}.
Functions are invoked via \emph{triggers} (\circledColor{A}{black}), including internal cloud events such
as database update or a new entry in a queue, and the standard external trigger via a cloud HTTP gateway
that exposes functions to the outside world.
A function scheduler (\circledColor{B}{black}) places the invocation in a cloud-native
execution environment (\circledColor{C}{black}),
and the function code is downloaded from the cloud storage (\circledColor{D}{black}).
Function are allowed to initiate connections to external
cloud resources and services, and can also use the filesystem of its sandbox as a temporary storage.
A sandbox instance handles many consecutive invocations, so
 resources are cached and reused across executions. 
\else
\noindent \textbf{Platform}
We characterize the FaaS platforms with a high-level overview presented in Fig.~\ref{fig:faas_model}
and refer interested readers to a wider
discussion in the literature~\cite{DBLP:journals/corr/abs-1902-03383,copik2021sebs,10.5555/3277355.3277369}.
Functions are invoked via \emph{triggers} (\circledColor{A}{black}), including internal cloud events such
as database update or a new entry in a queue, and the standard external trigger via a cloud HTTP gateway
that exposes functions to the outside world.
A function scheduler (\circledColor{B}{black}) places the invocation in a cloud-native
execution environment (\circledColor{C}{black}),
and the function code is downloaded from the cloud storage (\circledColor{D}{black}).
Function are allowed to initiate connections to external
cloud resources and services, and can also use the filesystem of its sandbox as a temporary storage.
A sandbox instance handles many consecutive invocations, so resources are cached and reused across executions. 
\fi

%
\iftr
\paragraph{Invocations}
The primary types are \emph{cold} and \emph{warm}.
\emph{Cold} invocations occur if the FaaS manager cannot find an idle sandbox for a given function,
and must allocate a new one. The latency includes the time 
In a \emph{warm} invocation, the function payload is sent
directly to the executing process.

The unpredictable and high-latency cold startups are a major
issue with serverless~\cite{coldStart,10.1145/3423211.3425682} as they can add seconds of overhead
to each invocation.
Modern lightweight virtual machines are designed to support low-latency and burstable
serverless invocations.
However, even warm invocations can incur significant overheads. On AWS Lambda, each invocation
is processed by a dedicated management service to decide function placement~\cite{246288}.
The function input is limited to a few megabytes, so users must transmit
larger payloads via the high-latency cloud storage.
The invocation's critical path is even longer in OpenWhisk~\cite{openwhisk}, as it includes a controller,
database, load balancer, and a message bus~\cite{sciabarra2019learning}.

In AWS Lambda, the RTT latency changes from 19.5 ms on 1kB to over 600 ms on 5MB,
and it varies between 30 ms and 75 ms on the size range typical for images passed to ML recognition functions (Fig.~\ref{fig:evaluation_faas}).
Since routing and allocation take at most 10 ms in warm invocations~\cite{246288},
the latency is dominated by network transmission.
Following Amdahl's law, reducing copying overheads and efficiently utilizing the fast in-cloud network
is the best opportunity to decrease serverless invocations costs by orders of magnitude.
\else
\noindent \textbf{Invocations}
\emph{Cold} invocations occur when no idle sandbox is available for a given function,
and must allocate a new one.
The latency includes an allocation of a new one, downloading the function code from external storage,
and starting an executor process.
In a \emph{warm} invocation, the function payload is sent directly to the executing process.
While lightweight virtual machines are designed to support burstable serverless invocations,
even warm invocations can incur significant overheads.
On AWS Lambda, each invocation
is processed by a dedicated management service to decide function placement~\cite{246288}.
The function input is limited to a few megabytes, so users must transmit
larger payloads via the high-latency cloud storage.
The critical path is even longer in OpenWhisk, as it includes a controller,
database, load balancer, and a message bus~\cite{sciabarra2019learning}.
Even though platforms optimize functions to execute in the same set of warm containers,
each invocation includes the repetitive placement logic of control plane.

In AWS Lambda, the RTT latency changes from 19.5 ms on 1kB to over 600 ms on 5MB,
and it varies between 30 ms and 75 ms on the size range typical for images passed to ML recognition functions (Fig.~\ref{fig:evaluation_faas}).
Since routing and allocation takes at most 10 ms in warm invocations~\cite{246288},
the latency is dominated by network transmission.
Following Amdahl's law, utilizing the fast network
is the best opportunity to decrease serverless invocations costs by orders of magnitude.
\fi

\iftr
\noindent \textbf{Observation \#3:} The multi-step invocation path is a barrier to achieving
zero-copy and fast serverless acceleration.
\toolname{} removes the centralized cloud proxies from invocations.
\else
\begin{summaryblue}{Observation}
\noindent The multi-step invocation path is a barrier to achieving
zero-copy and fast serverless acceleration.
\toolname{} removes the centralized cloud proxies from invocations.
\end{summaryblue}
\fi

\noindent \textbf{High-Performance Serverless}
\iftr
The elastic parallelism of FaaS has gained minor traction so far in the world of
high-performance and scientific computing~\cite{9284321},
but serverless is used in compute-intensive workloads
such as data analytics, video encoding, linear algebra, and machine learning~\cite{227653,Mller2019LambadaID,
DBLP:journals/corr/abs-1907-11465,234886,DBLP:journals/corr/abs-1810-09679,10.1145/3357223.3362711,8360337,8457817}.
Such applications require low-latency communication and optimized data movement.
They cannot tolerate the large overheads of invoking remote functions.
They need a pricing model that is fair towards compute-intensive
functions~\cite{DBLP:journals/corr/abs-1902-03383}.
Although recent research improved serverless performance by including RPC~\cite{nightcore},
exploiting data locality,
and co-locating invocations~\cite{10.5555/3277355.3277444},
latency-sensitive and parallel applications need fast remote invocations to achieve high scalability.
\noindent \textbf{Observation \#4:} Connection latency and bandwidth are the fundamental bottlenecks
for remote invocations, yet serverless platforms do not take advantage of modern network protocols.
\toolname{} integrates high-speed RDMA connections.
\else
While the elastic parallelism of FaaS has been used in compute-intensive workloads such as
data analytics, video encoding, and machine learning training~\cite{jiang2021towards,Mller2019LambadaID,10.5555/3154630.3154660},
it has only gained minor traction so far in
high-performance and scientific computing~\cite{9284321}
due to a lack of low-latency communication and optimized data movement.
Although recent research improved serverless performance by including RPC~\cite{nightcore},
exploiting data locality,
and co-locating invocations~\cite{10.5555/3277355.3277444},
latency-sensitive and parallel applications need fast remote invocations to achieve high
scalability.
\begin{summaryblue}{Observation}
\noindent Connection latency and bandwidth are the fundamental bottlenecks
for remote invocations, yet serverless platforms do not take advantage of modern network protocols.
\toolname{} integrates high-speed RDMA connections.
\end{summaryblue}
\fi

\subsection{Remote Direct Memory Access} 
\label{sec:background_rdma}
\iftr
RDMA-capable networks have become a standard tool for implementing
high-performance communication libraries, transactions, distributed protocols, storage, and databases
~\cite{10.1145/3127479.3128609,10.1145/2749246.2749267,10.1145/2815400.2815419,10.1145/2882903.2882949,
10.1145/3448016.3452817,6468497,cao2021polardb}.
Unlike in the standard TCP/IP stack, RDMA data transfers are performed entirely by a dedicated
network controller bypassing both the CPU and operating system.
Instead of exchanging OS-managed and buffered packets or datagrams, RDMA allows the communicating
endpoints to directly read, write and atomically update memory contents of its remote counterpart.
The CPU and cache hierarchy do not participate in this process, and the remote network controller
forwards the arriving data over the PCI bus to the memory and avoids latency introduced by the kernel.
Thus, the receiver is passive and potentially even unaware of the communication.

This communication protocol provides high-speed and rapid access to other server's data with a lower
CPU utilization at the cost of a simplified and crude interface.
Achieving the best performance requires fine-tuning such as aligning memory, controlling device
buffers, and utilizing vendor-specific optimizations, e.g., message inlining.
Error-handling is solely the programmer's responsibility, data transfers are restricted to
\emph{locked} memory pages, and security concerns might require additional mitigation
mechanisms~\cite{rothenberger2021redmark,254436}.
The RDMA devices must be accessed in cloud environments
through virtualization solutions such as PCI passthrough, para-virtualization, and
virtual device functions~\cite{MAUCH20131408}.
Finally, RDMA can be used with InfiniBand, Cray and Intel networks,
RDMA over Converged Ethernet (RoCE),
and software virtualizations~\cite{225978,softroce}.

\noindent \textbf{Observation \#5:} Cloud applications take advantage of low overhead and
high performance of RDMA networks, but FaaS needs a complete redesign to benefit from them.
The design of \toolname{} is RDMA-compatible from the start.

\else
RDMA-capable networks have become a standard tool for implementing
high-performance communication libraries, distributed protocols, storage, and databases.
Unlike in the TCP/IP stack, RDMA transfers are performed entirely by a dedicated
network controller bypassing both the CPU and operating system.
Data is forwarded over the PCI bus to the memory, allowing the communicating
endpoints to directly read, write and atomically update memory of its remote counterpart.
This communication protocol provides high-speed and rapid access to other server's data with a lower
CPU utilization at the cost of a simplified and crude interface.
Error-handling is solely the programmer's responsibility and achieving the best performance
requires fine-tuning such as aligning memory, controlling device
buffers, locking memory, and utilizing vendor-specific optimizations.
RDMA devices are accessed in multi-tenant environments through PCI passthrough, para-virtualization,
and virtual device functions~\cite{MAUCH20131408}.

\begin{summaryblue}{Observation}
Cloud and HPC applications take advantage of low overhead and high performance of RDMA networks.
\toolname{} adapts FaaS computing to be RDMA-compatible.
\end{summaryblue}
\fi


\section{RDMA-based Serverless Platform}

\label{sec:design}
\sloppy \toolname{} tailors serverless architectures to the needs of high-performance applications.
In \toolname{}, we combine the best of two worlds - resource flexibility offered by FaaS computing
with the low overhead communication primarily available in the cloud IaaS resources and HPC clusters.
\toolname{} improves the central FaaS paradigm of remote executions by replacing the REST and RPC
invocations with direct memory operations on remote servers.
The enhanced architecture provides the same semantics of executing user code on
ephemeral workers while avoiding the major performance overheads of serverless.
%
%

\begin{figure}[t]
  \centering
  \includegraphics[width=1.0\linewidth]{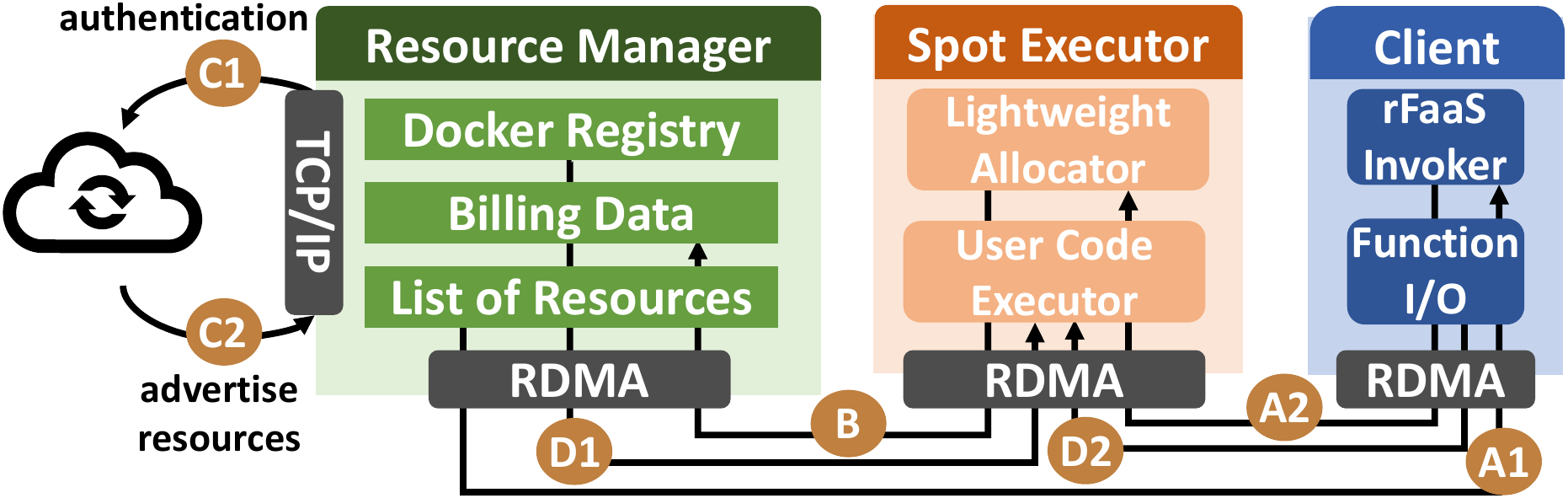}
  \caption{
    \textbf{
    \toolname{}: resource manager interacts with data center resources
    (\circledColorSmall{C}{brown}) and manages billing (\circledColorSmall{B}{brown}),
    clients acquire FaaS leases (\circledColorSmall{A}{brown}),
    and invoke functions (\circledColorSmall{D}{brown}).
  }
  }
  \vspace{-1.5em}
  \label{fig:system_design}
\end{figure}

Our philosophy in implementing \toolname{} is to drastically reduce the critical path of invocations.
We achieve this goal by reducing the number of parties involved in transmitting function data and
removing the centralized gateway and resource manager from the invocation path.
Compared to other architectures (Sec.~\ref{sec:background_faas}), we limit resource allocation and
authorization to cold startups, and remove both message queues and the bus for all warm invocations. 
First, we introduce \textbf{leases} to optimize the repeated allocation logic of FaaS control plane
(Sec.~\ref{sec:decentralized_allocation}).
%
%
Instead, our functions gain a direct RDMA connection to the user code executor without
sacrificing their \emph{serverless} nature (Fig.~\ref{fig:system_design}).
As in other FaaS platforms, no specific assumptions
about the underlying system and hardware are made.
We capitalize on this gain further by implementing an RDMA-based invocation
designed to minimize latency (Sec.~\ref{sec:design_invocation})
and handle parallel executions (Sec.~\ref{sec:design_scalability}).

%

\subsection{Components of \toolname{}}
\noindent \textbf{Resource Manager}
\toolname{} optimizes the FaaS control plane by splitting allocation and invocation
to avoid repeated function placement in the same small group of containers.
Instead, clients request and receive leases on spot executors (\circledColorSmall{A1}{brown}).
Cluster operators add and remove idle resources to the manager (\circledColorSmall{C2}{brown}),
and each instance of resource manager is responsible for a subset of spot executors.
With the lease concept, managers achieve the same load balancing and oversubscription targets as in
other platforms, while keeping the overhead low - they are not involved in warm and hot invocations,
which constitute the majority of serverless operations.
Mangers use heartbeats to verify the status of spot executor, and announce to clients
the lease termination to support fast resource reclamation.

\noindent \textbf{Spot Executor}
When clients begin offloading tasks to \toolname{}, they acquire leases on spot executors
to achieve the desired number of parallel workers.
These servers offer idle and unused hardware resources (CPU cores, memory) to support the
dynamic execution of serverless functions.
Clients connect to the lightweight allocator
(\circledColorSmall{A2}{brown}),
which is responsible for connecting new clients, managing
user code executors, removing processes that are idle for a long time or exceed specified
time limits, and accounting for resource consumption.
The allocator initializes an isolated execution context with an RDMA-capable execution process.
Finally, clients can establish a direct RDMA connection with each
executor process and invoke functions by writing function header and
payload directly into their memory (\circledColorSmall{D2}{brown}).
The results are returned to the client in a similar fashion, and the client caches the lease
for consecutive invocations on warmed-up resources.


\subsection{Allocation Leases}
\label{sec:decentralized_allocation}
%
Decentralized resource management in form of leases is another improvement that \toolname{}
brings to serverless.
To execute a function, clients involve the resource manager only once to acquire leases
(\circledColorSmall{A1}{brown}).
%
%
%
%
Clients cache connections to executor processes and use them for consecutive executions on warmed-up resources,
helping to avoid the initialization costs for reliable RDMA connections.
To support straightforward deallocation of on-demand executors,
clients use the connection status to check if the process is alive.
When users terminate the allocation before the lease expires, executors notify the manager
to include their resources in future allocations.

\begin{figure}[t]
  \centering
  \includegraphics[width=1.0\linewidth]{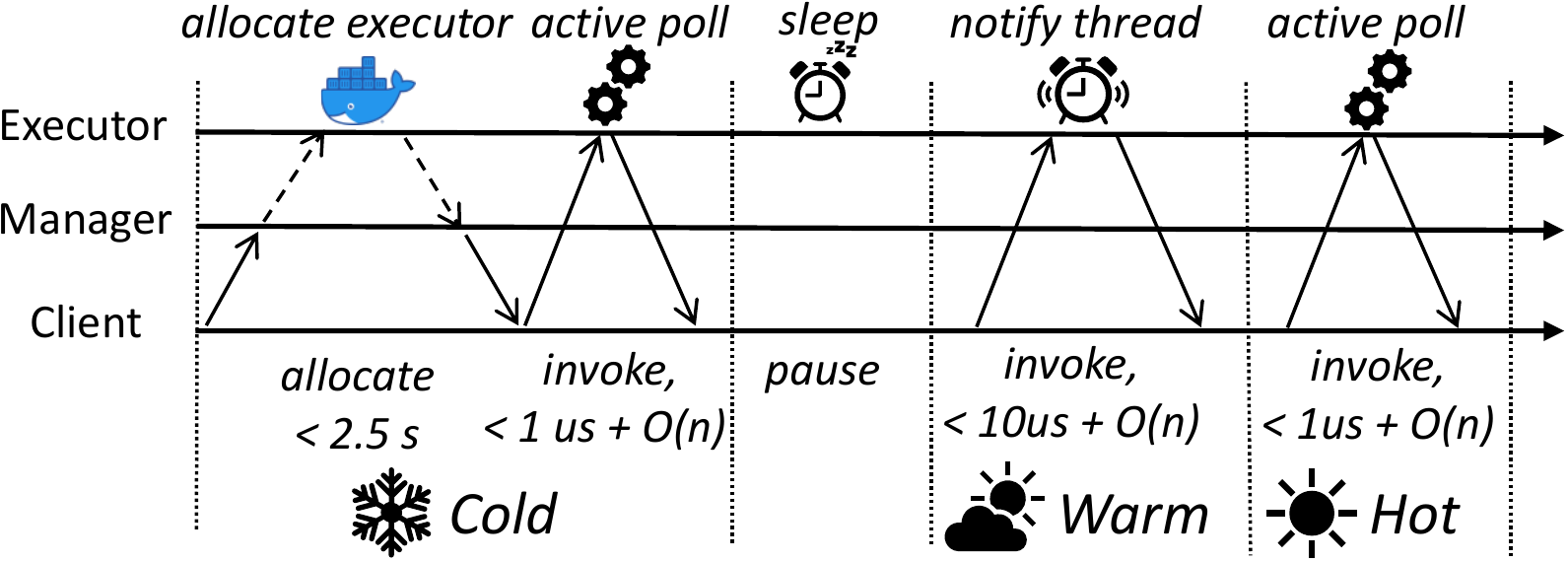}
  \caption{
    \textbf{Lifetime of a function in \toolname. Cold start times
  are dominated by sandbox initialization. Warm and hot invocation times include \toolname overhead
  and latency of RDMA write of $N$ bytes of payload.}
}
  \vspace{-2em}
  \label{fig:function_invocation}
\end{figure}

\subsection{Low-Latency Invocation}
\label{sec:design_invocation}
%
A critical feature of \toolname{} is ensuring invocations have the lowest overhead possible.
While an on-demand allocation of idle resources improves the economics of the data canter,
it would be counterproductive to incorporate \toolname{} functions into high-performance
applications if we did not offer fast invocations.
%
%
In addition to standard FaaS \emph{cold} and \emph{warm} invocations, we provide
a new \emph{hot} invocation type that guarantees zero-copy execution on
pre-allocated hardware (Fig.~\ref{fig:function_invocation}).

\noindent \textbf{Cold}
The \emph{cold} invocation includes significant overheads caused by the initialization of an execution context.
In \toolname{}, clients acquire leases by
requesting the desired core count, memory, and timeout for the allocation.
%
%
Then, the lightweight allocator initializes an isolated execution sandbox and assigns the
requested computing and memory resources to it.
The user code executor starts in the sandbox, accesses the selected RDMA device, registers
memory buffers, and creates worker threads pinned to assigned cores.
Each executor process has a configurable number of thread workers who work independently,
and each one corresponds to a single function instance.
\iftr
Thus, clients can allocate multiple workers in a single allocation request.
\fi
When the initialization is done, the client receives connection settings,
establishes connections to all threads, and invokes functions by
writing requests directly to the workers' remote memory.
%
%
Overall, sandbox initialization adds on average 25 ms and 2.7 seconds of overhead for bare-metal
and Docker-based executors, respectively, on an HPC node (Sec.~\ref{sec:evaluation}).
%

\noindent \textbf{Warm}
%
The client transmits the function payload using an RDMA connection to an allocated executor.
Executor threads do not share RDMA resources,
and they use blocking wait independently to receive completion events corresponding
to new \emph{warm} invocation requests.
%
%
Using blocking wait increases latency
but significantly decreases the pressure on computing resources compared to active polling.
In the unlikely case of resource exhaustion on the executor, the invocation request is immediately rejected
and the client redirects the invocation to another executor (Fig.~\ref{fig:invocation-workflow}).
Compared to native RDMA performance of a round-trip communication, warm invocations have an
overall overhead of fewer than 6 microseconds.

\noindent \textbf{Hot}
The novel \emph{hot} invocation improves the performance of \emph{warm} FaaS executions
by adding the obligation that threads actively poll for invocation requests.
%
%
The busy polling decreases the invocation latency since threads do not enter
a blocked state to wait for an interrupt generated by the RDMA driver.
The thread enters the \emph{hot} invocation mode immediately after execution and polls RDMA events
without sleeping to improve the performance of consecutive invocations.
Executors can roll back to \emph{warm} executions to free up the CPU after a configurable time without a new invocation,
depending on user preferences.
%
This configuration decreases the overall overhead for a round-trip invocation
to ca. 300 nanoseconds on average.
However, it comes at the cost of occupying the CPU core and preventing other functions
from using the computing resources.
Therefore, the \emph{hot} polling time should be accounted as active computation time.
In exchange, users gain always available computing resources, helping to incorporate
functions into HPC applications and support iterative invocations.

\subsection{Scalability}
\label{sec:design_scalability}
A high-performance serverless platform must handle scaling in three directions:
number of spot executors,
number of \toolname{} users,
and the number of functions invoked by a client.
\noindent \textbf{Horizontal Scaling}
The number of spot executors and clients in \toolname{} is bounded by the size of the RDMA network.
\iftr
While modern cloud RDMA networks count many thousands of clients~\cite{8891004,guo2016rdma}
the networks can scale globally in future deployments.
Since the network throughput of RDMA connections decreases
significantly with the number of clients~\cite{10.1145/3302424.3303968,anujKVS},
the resource manager is replicated like in other FaaS platforms.
\else
Since the network throughput of RDMA connections decreases
significantly with the number of clients~\cite{anujKVS},
the resource manager is replicated like in other FaaS platforms.
While modern RDMA networks and supercomputers count many thousands of clients~\cite{guo2016rdma}
the networks can scale globally in future cloud deployments.
\fi
Resources are split between manager instances and round-robin scheduling allows handling
the increasing number of lease requests from clients.
%
%
%
%


\noindent \textbf{Parallel Invocations}
\toolname{} allows for simultaneously dispatching
function execution requests to threads of remote user executors.
The user requests how many function instances should be used, and the client library manages lease
allocations to reach the desired scale.
The client has a direct RDMA connection to each thread worker and can invoke functions concurrently.
Function workers operating on the same node are independent of each other
and can all execute different functions.
The scalability is achieved by exploiting the non-blocking nature of RDMA write operations
and using disjoint memory buffers to store results. 
Multiple RDMA connections improve network utilization as
more processing units of a network controller are involved~\cite{anujKVS}. 
%
%
Each executor thread
switches between hot and warm
invocations on its own, further aiding elasticity.

\begin{figure}[t]
  \centering
  \includegraphics[width=1.0\linewidth]{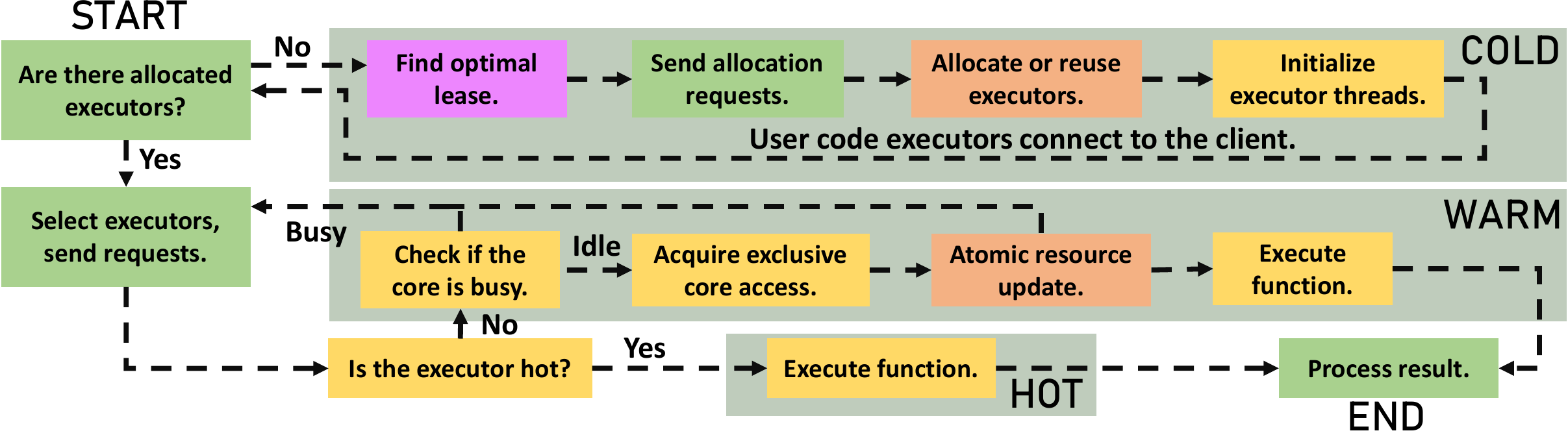}
  \vspace{-1.5em}
  \caption{
    \textbf{
      \toolname{} invocations include RDMA communication between clients (\circledColorX{newgreen}),
  resource manager (\circledColorX{newpurple}), and spot executors (\circledColorX{neworange}) with user code executors (\circledColorX{newyellow}).
    }
  }
  \vspace{-1em}
  \label{fig:invocation-workflow}
\end{figure}

\ifcnf

\noindent \textbf{Oversubscription}
FaaS platforms oversubscribe resources since invocations often arrive independently, at different times,
and consume different resources.
This aligns well with the environments of scientific clusters, where large amounts of free memory
can be used to retain more warm sandboxes than available CPU cores.
However, many HPC applications cannot tolerate the overhead and imbalance introduced by
oversubscribed execution, even if such an event is unlikely.

To that end, \emph{hot} invocations ensure that the executor occupies the CPU core and handles
the request immediately (Fig.~\ref{fig:invocation-workflow}).
Since this guarantee is not always needed and can be expensive. \emph{warm} invocations are
executed opportunistically on resources that might be oversubscribed.
A successful warm invocation requires only a single, local RDMA communication between spot executors
and its allocator to verify resource status, thus the additional latency is negligible for
larger function payloads.
When resources are unavailable, the request is rejected, and the client sends it again to
another user code executor.
Thanks to the RDMA networking, the rejection can be processed with microsecond latency,
minimizing the performance hit.
%
%
Warm invocations avoid interference with compute-intensive tasks on the same CPU core since
rejection is a short and I/O-intensive process.
%
Hot and warm invocations can be switched dynamically, providing the performance and flexibility needed
for all types of HPC computations.

\subsection{Isolation and Security}
\toolname{} aims to provide the same security level as serverless invocations in the cloud.
Multi-tenant environments require that functions execute in isolation, and the user's code is prohibited
from accessing any resources, data, and code not provided with the invocation.
Thus, in addition to bare-metal executors, we include containerized executors.
The main requirements imposed by \toolname{} are virtualization support for RDMA-capable network
controllers and negligible performance overheads.
%
%
\toolname{} uses Docker containers to implement isolated execution contexts for user functions.
Single Root I/O Virtualization (SR-IOV) provides high-performance virtualized network controllers
in a multi-tenant environment~\cite{5416637}.
%
%
On platforms without SR-IOV, we can use software virtualization systems such as FreeFlow~\cite{225978}.

\toolname{} leases shift the control plane involvement from each execution to cold invocations,
but they do not introduce additional security challenges.
Leases are time--limited and include user authentication.
Thus, they are similar to batch system allocations
that release resources to a job and perform authorization only once.
Furthermore, modern RDMA extensions provide authentication, payload encryption, and memory protection
ensuring secure transmission in multi-tenant networks~\cite{254436,10.1145/3387514.3405897,ipsecroce}.

\subsection{Modularity}
The world of high-performance applications and cloud systems is rich and diverse.
Thanks to its modular design, \toolname{} supports extensions into new environments and hardware.

\paragraph{Network}
While our implementation manages RDMA networks with \code{ibverbs},
the \toolname{} functionality is orthogonal to the device interface
and can be implemented with higher-level concepts from \code{libfabric}~\cite{libfabric}.
\toolname{} can be deployed on other networks providing RDMA-like semantics,
such as the Elastic Fabric Adapter in the AWS cloud~\cite{awsefa}.
In addition, software virtualization can be employed in data centers without high-speed networks,
offering RDMA semantics at the cost of
higher overheads~\cite{softroce,225978}.

\paragraph{Language}
\toolname{} supports C and C++ functions and native integration into C/C++ applications (Sec.~\ref{sec:integration}).
The language choice is, however, independent from the platform itself.
\toolname{} functions can be implemented effortlessly in
languages ABI-compatible with C, such as Rust,
and with the help of foreign-function interface in languages prevalent in the serverless community, such as Python.

\paragraph{Sandbox}
\toolname{} functions can be served in other environments than bare-metal processes or Docker containers,
e.g., in HPC container Singularity~\cite{10.1371/journal.pone.0177459},
gVisor~\cite{234857},
and in microVMs such as Firecracker~\cite{firecracker,246288}
that provide a higher level of isolation with negligible performance overheads.
New sandbox types can be integrated effortlessly as long as a virtualization
or passthrough to the RDMA NIC is provided.

\fi

\iftr
\subsection{Resource Exclusivity}
Many FaaS platforms allow oversubscribing resources since invocations often arrive independently, at different times, and have compatible
resource consumption.
Servers can host many more warm sandboxes than available CPU cores since this resource is limited primarily by the memory capacity.
This feature aligns well with the environments of scientific clusters, where large amounts of free memory can be used to keep
warm containers.
However, high-performance and compute-intensive applications cannot tolerate the overhead and imbalance introduced when two invocations arrive at the same
on the same core, even if such an event is unlikely.

Thus, the \emph{hot} invocation provides the performance guarantee for consecutive and streamlined invocations, where the function
executor occupies the CPU core and is immediately ready to handle requests (Fig.~\ref{fig:invocation-workflow}).
However, this guarantee is not always needed, and it comes with additional costs.
Instead, users can send \emph{warm} invocation requests to sandboxes that can fail in the case when actual oversubscription happens.
Hot and warm invocations can be switched dynamically, providing the performance and flexibility needed for all types of computation offloading in HPC applications.
When resources are unavailable, the request is rejected, and the client sends it again to another user code executor. 
Thanks to the RDMA networking, the rejection can be processed with microsecond latency, minimizing the performance hit.
Warm invocations are implemented to avoid interference with compute-intensive tasks that can occupy the same CPU core.
The rejection is an I/O-intensive process that takes a very short time, and its requirements for CPU time are further decreased when simultaneous multithreading is available.
A successful warm invocation requires only a single, local RDMA communication with the spot executor allocator, and the additional latency is negligible for larger function payloads.

\subsection{Fault tolerance}
\toolname{} clients can experience failures in three ways:
the spot executor is evicted,
the executor shuts down uncontrollably,
or the function crashes the execution process.
Even though the likelihood of a failure caused by the eviction of harvested cloud resources is very low ($< 0.002\%$)~\cite{10.1145/3477132.3483580},
we provide mitigation strategies for a seamless user experience.
Executor connections are cached in the client library for further invocations, and a server failure is detected through a disrupted RDMA connection.
The library repeats the invocation on other servers
for a finite number of retries to avoid an infinite loop on a broken function.
Furthermore, the spot executor manager frequently verifies the status of its executors,
and it notifies the client when it detects a premature exit and failure.

\subsection{Isolation and Security}
Multi-tenant environments require that functions execute in isolated sandboxes with restricted access to system resources.
Thus, in addition to bare-metal executors, we include containerized executors to ensure
privacy and security in the multi-tenant execution in \toolname{}.
The main requirements imposed by \toolname{} are virtualization support for RDMA-capable network
controllers and negligible performance overheads.
The current implementation uses Docker containers to implement isolated execution contexts for user functions.
%
The container prohibits the user's code from accessing any resources, data, and code not provided with the
invocation.
We use the Single Root I/O Virtualization (SR-IOV) to virtualize network controllers
in a multi-tenant environment.
Virtual network functions provide isolated but
high-performance access for different users~\cite{5416637}.
\toolname{} managers use a set of pre-configured virtual device functions to deploy containers.
rFaaS is compatible with other sandbox types common in the serverless stack,
such as microVMs and unikernels~\cite{246288,234857},
as long as a passthrough or virtualization access to the rNIC is provided.
On platforms without SR-IOV, we can use software virtualization systems such as FreeFlow~\cite{225978}.

\toolname{} aims to provide the same level of security in hot and warm invocations as in regular
FaaS invocations in the cloud.
We are unaware of any negative security implications when removing control plane involvement in each invocation in a sequence.
\toolname{} leases are time--limited and include user authentication.
Thus, they are similar to batch system allocations
that release resources to a job and perform the authorization only once.
Furthermore, modern RDMA extensions provide authentication, payload encryption and memory protection
that ensure secure transmission in cloud networks~\cite{254436,10.1145/3387514.3405897,ipsecroce}.

\subsection{Modularity}
The world of high-performance applications and cloud systems is rich and diverse.
Thanks to its modular design, \toolname{} supports extensions into new environments and hardware.

\paragraph{Network}
While our implementation manages RDMA networks with \code{ibverbs},
the \toolname{} functionality is orthogonal to the device interface
and can be implemented with higher-level concepts from \code{libfabric}~\cite{libfabric}.
\toolname{} can be deployed on other networks providing RDMA-like semantics,
such as the Elastic Fabric Adapter in the AWS cloud~\cite{awsefa}.
In addition, software virtualization can be employed in data centers without high-speed networks,
offering RDMA semantics at the cost of
higher overheads~\cite{softroce,225978}.

\paragraph{Language}
\toolname{} supports C and C++ functions and native integration into C/C++ applications (Sec.~\ref{sec:integration}).
The language choice is, however, independent from the platform itself.
\toolname{} functions can be implemented effortlessly in
languages ABI-compatible with C, such as Rust,
and with the help of foreign-function interface in languages prevalent in the serverless community, such as Python.

\paragraph{Sandbox}
\toolname{} functions can be served in other environments than bare-metal processes or Docker containers,
e.g., in HPC container Singularity~\cite{10.1371/journal.pone.0177459},
gVisor~\cite{234857},
and in microVMs such as Firecracker~\cite{firecracker,246288}
that provide a higher level of isolation with negligible performance overheads.
New sandbox types can be integrated effortlessly as long as a para-virtualization
or passthrough to the RDMA NIC is provided.

\subsection{\toolname{} versus \emph{traditional} FaaS}
\label{sec:design_versus}
%
\toolname{} is a fundamental building block for bringing RDMA abstractions into serverless.
While \toolname{} implements the essential semantics of FaaS computing - remote invocations
on transient and multi-tenant resources with pay-as-you-go billing -
we tailor the design of the serverless platform to the demands of high-performance and low-latency.
%
%
On top of the standardized interface of \toolname{}, additional features can be implemented
according to the needs of specific programming frameworks: native datatypes and serialization,
collective operations, logging of invocations.
Similarly, \toolname{} does not come with a dedicated authorization system, as that can be provided
through the existing cloud solutions.

\paragraph{Triggers}
The trigger mechanism is replaced with decentralized allocations and direct connection to
an executor process, removing the proxies and caching the connection
to minimize latency and unnecessary copies.
However, our RDMA abstractions are compatible with the serverless execution model and can be used to provide centralization
known from other FaaS systems at the cost of higher overheads.
For example, such a platform could utilize RDMA-aware queues and storage~\cite{10.1145/3148055.3148068,balasubramanianfastps}
to implement a default entry point for function triggering.
Users can still achieve orders of magnitude improvement in invocations latencies
with the end-to-end latency of as low as 100 microseconds in RDMA Kafka~\cite{kafkadirect}
and single-digit microseconds for RDMA native publish-subscribe service~\cite{balasubramanianfastps}.
Furthermore, centralized FaaS gateways can benefit from RDMA acceleration in the cloud network by using
\toolname{} abstractions in the backend, preserving a consistent interface for users.
\fi

\section{\lowercase{r}F\lowercase{aa}S in Detail}

\label{sec:integration}


\setlength{\textfloatsep}{0pt}
\begin{listing}
\begin{minted}[escapeinside=||,fontsize=\footnotesize]{c++}
uint32_t f(void* in, uint32_t size, void* out) {
 uint32_t in_len = size / sizeof(double);
 double* input = reinterpret_cast<double*>(in);
 double* output = reinterpret_cast<double*>(out);
 // Run function's code.
 uint32_t out_len = solve(input, in_len, output);
 // Return value defines the output size
 return sizeof(double) * out_len;
}
\end{minted}
\vspace{-0.1em}
\caption{rFaaS function interface.}
\label{lst:example}
\end{listing}

\iftr
\toolname{} functions are aligned with existing cloud frameworks
(Fig.~\ref{fig:system_design}, p.~\pageref{fig:system_design}).
Functions are deployed (Sec.~\ref{sec:function_deployment}) and invoked with a user-oriented
and high-level C++ programming interface (Sec.~\ref{sec:integration_programming}).
\else
\toolname{} functions are deployed as containers (Sec.~\ref{sec:function_deployment}).
We improve over the HTTP-based REST interfaces in other FaaS platforms,
and provide a user--oriented C++ interface to improve performance and hide RDMA complexity
(Sec.~\ref{sec:integration_programming}).
\fi
%
\subsection{Function Deployment}
\label{sec:function_deployment}

Listing~\ref{lst:example} presents the standard function interface in \toolname{}.
%
%
Input is written to memory buffers of the user code executor.
while the RDMA immediate value contains an invocation identifier and a function index. 
%
%
The function returns the number of bytes in the output array sent back to the client.
The input buffer contains a twelve-byte \emph{header} with an address and access key
for a buffer on the client's side, and the executor writes the output directly to
the client's memory.
%

\toolname{} supports the execution of arbitrary functions,
and similarly to the \emph{function app} offered by Azure Functions,
we enable the execution of different functions in the same worker process.
%
%
%
\iftr
We offer users two ways of distributing their serverless functions: code package
and Docker images.
Users upload a package with the function's code to the storage service in the first option.
Alternatively, users can deploy a Docker image containing the function's code and dependencies.
The image is enriched with \toolname{} RDMA executor
and placed in a hosted Docker registry (\circledColorSmall{D1}{brown}).
\else
Serverless functions are deployed as containers with code and all dependencies.
%
%
%
The image is enriched with \toolname{} RDMA executor
and placed in a Docker registry.
\fi
%

%
%
%
%
%

\subsection{Programming Model}
\label{sec:integration_programming}

\begin{figure}[t]
	\centering
  \includegraphics[width=1.0\linewidth]{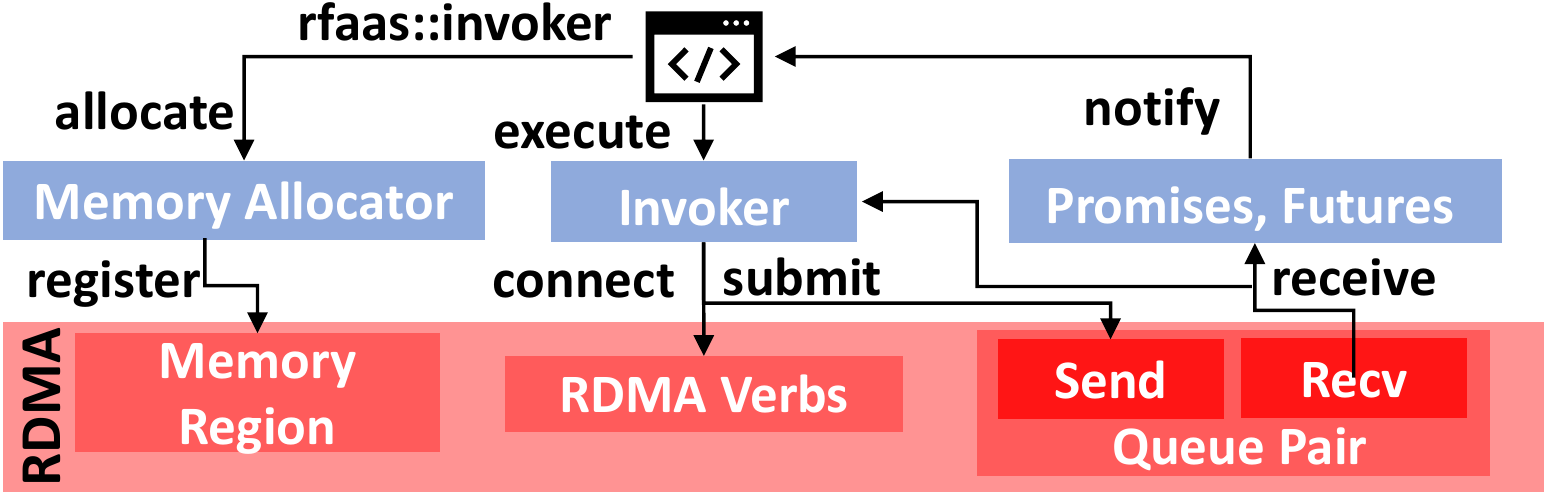}
  \caption{\textbf{The programming model of \toolname{}, inspired by C++ standarization
  efforts on the \emph{executor} concept.}}
  \label{fig:programming_model}
\end{figure}

\AtBeginEnvironment{minted}{\setlength{\parskip}{0pt}}
\AtEndEnvironment{minted}{\setlength{\parskip}{0pt}}
\begin{listing}
\begin{minted}[escapeinside=||,fontsize=\footnotesize]{c++}
void compute(int size, options & opts) {
 rfaas::invoker invoker{opts.rnic_device};
|\llap{\textrm{\ding{172}}}| invoker.allocate(opts.lib, opts.size * sizeof(double),
   rfaas::invoker::ALWAYS_WARM_INVOCATIONS);
|\llap{\textrm{\ding{173}}}| auto alloc = invoker.allocator<double>{};
 // Automatically expanded with function's header
|\llap{\textrm{\ding{174}}}| rfaas::buffer<double> in = alloc.input(2 * size);
 rfaas::buffer<double> out = alloc.output(2 * size);
 // Offload part of the computation to rFaaS
|\llap{\textrm{\ding{175}}}| auto f = invoker.submit("task", in, size, out);
 local_task(in.data() + size, out.data() + size, size);
|\llap{\textrm{\ding{176}}}| f.get();
|\llap{\textrm{\ding{177}}}| invoker.deallocate(); // Release computing resources.
}
\end{minted}
\caption{Example of an rFaaS-accelerated application.}
\label{lst:example_integration}
\end{listing}

To design the programming interface for \toolname{}, we take inspiration from recent
developments in the C++ standard for parallel and asynchronous \emph{executors}~\cite{P0443R14}.
The prior work on executors and their implementations proved that this concept
is an efficient interface for dispatching tasks to accelerator devices~\cite{10.1007/978-3-319-46079-6_2,10.1145/3078155.3078187}.
The programming model presented in Fig.~\ref{fig:programming_model} hides the complexity of RDMA
verbs under a lightweight C++ abstraction.
As a result, it can be easily integrated into existing parallel applications
as presented in Listing~\ref{lst:example_integration},
and it can be adapted in the future to full compatibility with C++.

\noindent \textbf{Memory Allocator}
The \emph{memory allocator} (\ding{173}) provides RDMA-enabled memory buffers
and encapsulates the memory region reserved for the function header (\ding{174}).
The allocator can be integrated effortlessly
to serialize standard C++ containers such as \emph{std::vector}
and \emph{std::array}, and all memory buffers are page-aligned to achieve
the highest bandwidth on RDMA~\cite{anujRDMA}.

\noindent \textbf{Invoker} 
\iftr
The client's \emph{invoker} implements the submission of remote function invocations (\ding{175}).
It also manages RDMA connections to remote executors and implements the allocation
and deallocation of computing resources.
The status of the computation can be queried with busy polling to minimize latency,
and we use the \code{std::future} to represent the result of unfinished executions.
Users can query the status of each invocation, wait for its completion, and access the result later (\ding{176}).
Internally, the library runs a single thread that waits for RDMA completion events
and modifies future's status when the corresponding invocation finishes.
%
The additional thread can be replaced with manual message progression.
While blocking wait has a higher latency than active polling~\cite{6332248},
the background thread sleeps, helping to reduce CPU consumption.

The allocation of \toolname{} functions can be performed ahead of time (\ding{172})
to hide the cold invocation latencies
since warm executor threads are sleeping and not incurring major charges.
Remote resources are allocated and deallocated as needed (\ding{177}),
adjusting to the varying parallelism and workload.
%
\else
The client's \emph{invoker} submits remote function invocations (\ding{175}).
It manages RDMA connections and implements the allocation and deallocation of leases.
The status of the computation can be queried with busy polling to minimize latency,
and we use the \code{std::future} to represent the result of unfinished executions.
Users can query the status of each invocation, wait for its completion, and access the result later (\ding{176}).
Internally, the library runs a single thread that waits for RDMA completion events
and modifies future's status when the corresponding invocation finishes.
While blocking wait has a higher latency than active polling~\cite{6332248},
the background thread sleeps, helping to reduce CPU consumption.

The allocation of \toolname{} functions can be performed ahead of time (\ding{172})
to hide the cold invocation latencies
since warm executor threads are sleeping and not incurring major charges.
Remote resources are allocated and deallocated as needed (\ding{177}),
adjusting to the varying parallelism and workload.
%

\setlength{\abovedisplayskip}{0pt}
\setlength{\belowdisplayskip}{0pt}
\subsection{Billing}
\label{sec:integration_billing}
\toolname{} uses a pricing model similar to provisioned serverless functions to include active
\emph{hot} polling.
The billing model includes three cost components: allocation time $C_{a}$, hot polling $C_{h}$,
and active computation time $C_{c}$.
\begin{equation*}
  C \: = \: C_{a} \cdot t_{a} + C_{c} \cdot t_{c}  + C_{h} \cdot t_{h}
\end{equation*}
The total allocation $t_{a}$ measured is calculated across all executors as a product of
allocation time and memory requested, whereas the active computation time $t_{c}$ and hot
polling time $t_{h}$, measured in seconds, represent the total time all remote workers were
busy with executing functions and polling for new invocations, respectively.
Thus, cluster operators can encourage warm invocations to boost utilization through resource overallocation,
while applications requiring the highest performance pay the premium for nanosecond invocation overheads.
The billing procedure is implemented in a global database associated with the resource manager
using RDMA atomic \emph{fetch-and-add} operations, providing lightweight allocators with an RDMA-native
way of accumulating cost results.

\fi

\iftr
\subsection{Authorization}
\label{sec:integration_authorization}

\toolname{} integrates with existing cloud authorization systems through a system of access tokens.
The cloud offers a user-facing API to generate access tokens and an internal API to verify
the token's authenticity.
The user passes the token to the allocator, who optimistically begins the allocation of user code executor
while verifying the token in the background with the help of the resource manager (\circledColorSmall{C1}{brown});
only the cloud-managed resource manager has permission to access the token verification system.
After a successful verification, the resources are released to the client, and the resource manager
shares remote billing buffer details with the lightweight allocator.
Thus, \toolname{} makes no assumptions on the provided authorization systems, the credentials
verification is conducted entirely on the cloud provider side, and
the RDMA token transmission can be encrypted and secured with modern systems~\cite{254436}.

\vspace{-1em}
\setlength{\abovedisplayskip}{0pt}
\setlength{\belowdisplayskip}{0pt}
\subsection{Billing}
\label{sec:integration_billing}
The pricing of \toolname{} is presented in the equation below, and it includes three basic
cost components: allocation of computing resources $C_{a}$, hot polling $C_{h}$,
and active computation time $C_{c}$.
\begin{equation*}
  C \: = \: C_{a} \cdot t_{a} + C_{c} \cdot t_{c}  + C_{h} \cdot t_{h}
\end{equation*}
The total allocation $t_{a}$ measured in GB-second is calculated
across all executors as a product of allocation time and memory requested.
Whereas the active computation time $t_{c}$ and hot polling time $t_{h}$,
measured in seconds, represent the total time all remote workers are busy with executing
functions and polling for new invocations, respectively.
Costs $C_{a}$, $C_{h}$ and $C_{c}$ are measured in \$ per GB-second and \$ per second
and represent the total cost of occupying and actively using system resources (i.e., cores, memory)
for a given period.
The pricing system of \toolname{} is similar to traditional FaaS systems
for provisioned function invocations~\cite{awsPricing},
where clients are charged for the pre-allocation of cloud resources.
However, unlike traditional FaaS, \toolname{} does not charge for invocation calls,
only for active CPU time.
As a result, clients of \toolname{} pay neither $C_{c}$ nor $C_{h}$
for times when remote executor threads sleep due to warm invocations. 
However, the cloud operators might increase the $C_{h}$ cost component to encourage
warm invocations, as their low CPU overhead helps further to boost utilization through
over-allocating resources on multi-tenant servers.
Applications requiring the highest performance can pay the premium for nanosecond
invocation overheads.

The billing procedure is implemented in a global database associated with the resource manager
(\circledColorSmall{B}{brown}).
The manager exposes memory regions for RDMA atomic \emph{fetch-and-add} operations, providing
lightweight allocators with an RDMA-native way of accumulating cost results without consuming CPU
resources.
%
%
We accumulate charges with a granularity of one second, and billing data is updated with
the same frequency.
%
This avoids a loss of accounting data due to the abrupt termination of \toolname{} spot executors.
The contention of atomic operations is not an issue, as cost accumulation is never on
the critical path of function invocation.
\fi

\section{\lowercase{r}F\lowercase{aa}S in Practice}
\label{sec:evaluation}
To demonstrate the fitness of \toolname{} for HPC,
we answer critical questions in the form of extensive evaluation.

\begin{enumerate}[noitemsep,nosep]
\item Is \toolname fast enough for high performance, latency-sensitive applications?
\item Are the overheads for initialization prohibitively large?
\item Does the \toolname{} bandwidth scale with larger payloads?
\item Does \toolname{} scale with more parallel workers?
\item Does \toolname{} integrate functions into HPC applications?
\item Is the performance of remote computing with \toolname{} competitive compared to local computation?
\item Are short \toolname{} functions usable in HPC computations?
\end{enumerate}

\noindent\textbf{Platform}
We deploy \toolname{} in a cluster and execute benchmark code on 4 nodes,
each with two 18-core Intel Xeon Gold 6154 CPU @ 3.00GHz and 377 GB of memory.
The nodes are equipped with a Mellanox MT27800 Family NIC with a 100 Gb/s Single-Port link configured with RoCEv2 support.
%
Nodes communicate with each other via a switch,
and we measured an RTT latency of 3.69 $\mu$s and a bandwidth of 11,686.4 MiB/s.
We use Docker 20.10.5 with the executor image \texttt{ubuntu:20.04},
and we use Mellanox's SR-IOV plugin to run containers over virtual device functions.
\toolname is implemented in C++, using g++ 8.3.1.

\subsection{Invocation Latency}
%




%
We begin by measuring the hot and warm invocation latency, using a no-op ''echo'' function that returns the provided input.
We use a warmed-up, single-threaded, bare-metal executor with the main thread pinned to a CPU core,
perform 10,000 repetitions, and report the median.
We compute the non-parametric 99\% confidence intervals of the median and find that the interval bounds are very tight ($<1\%$).
To assess the overheads of \toolname invocations, we measure the latency
of RDMA and TCP/IP transmissions. 
For the former, we report the median of the \emph{ib\_write\_lat} benchmark executed with thread pinning.
For the latter, we report the mean of \emph{netperf} used with page-aligned buffers and process pinning.

\begin{figure}[t]
	\centering
  \includegraphics[width=1.0\linewidth]{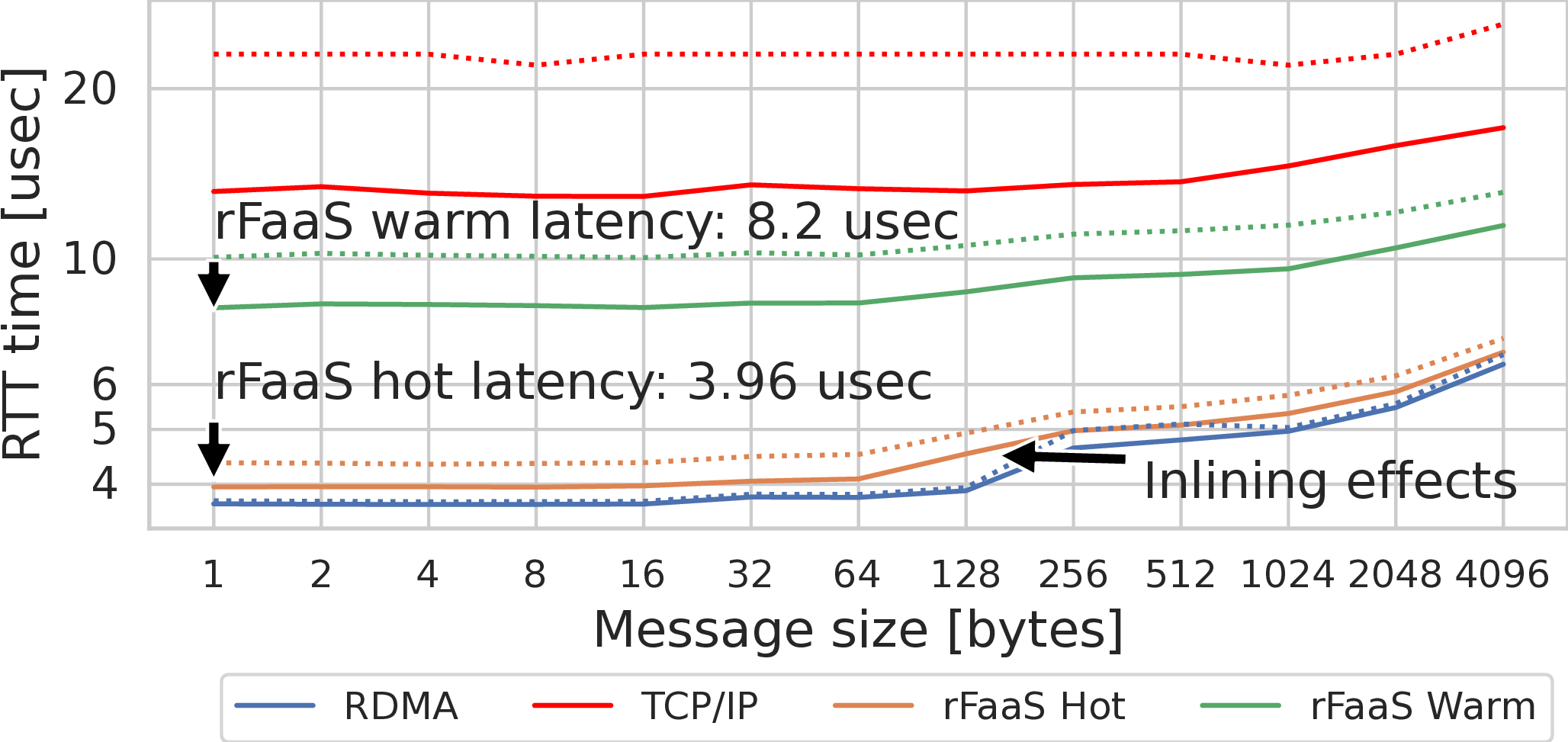}
  \caption{\textbf{The RTT of an no-op \toolname{} function and network transport,
  median (solid) and 99th latency (dashed).}}
  \label{fig:evaluation_hot_warm}
\end{figure}

Fig.~\ref{fig:evaluation_hot_warm} shows
that the overhead imposed by processing a no-op function by \toolname{} in a process is 326 ns on average, compared to the baseline RDMA data transmission.
The measurements for a Docker-based executor present additional ca. 50 ns overhead over RDMA writes
when using a container.
The only exception is the message size of 128 bytes, where the overhead increases to 630 ns.
There, RDMA can use message inlining for both directions of transmission to improve performance of small messages~\cite{anujKVS}.
However, the communication in \toolname is asymmetric: we transmit 12 more bytes for the input, forcing us to use non-inlined write operations in one direction.
The average overhead of a warm execution is 4.67 $\mu$s, and containerization adds a latency of ca. 650 ns. 

With slightly more than 300 ns of overhead, we enable remote invocations without noticeable performance penalty,
conclusively answering: \textbf{\toolname is fast enough for latency-sensitive, high-performance applications}.

\subsection{Cold Invocation Overheads}
\begin{figure}[tb!]
	\centering
  \resizebox*{0.95\width}{0.85\totalheight}{
    \subfloat[Bare-metal Linux process
    ]{%
      \includegraphics[width=1.0\linewidth]{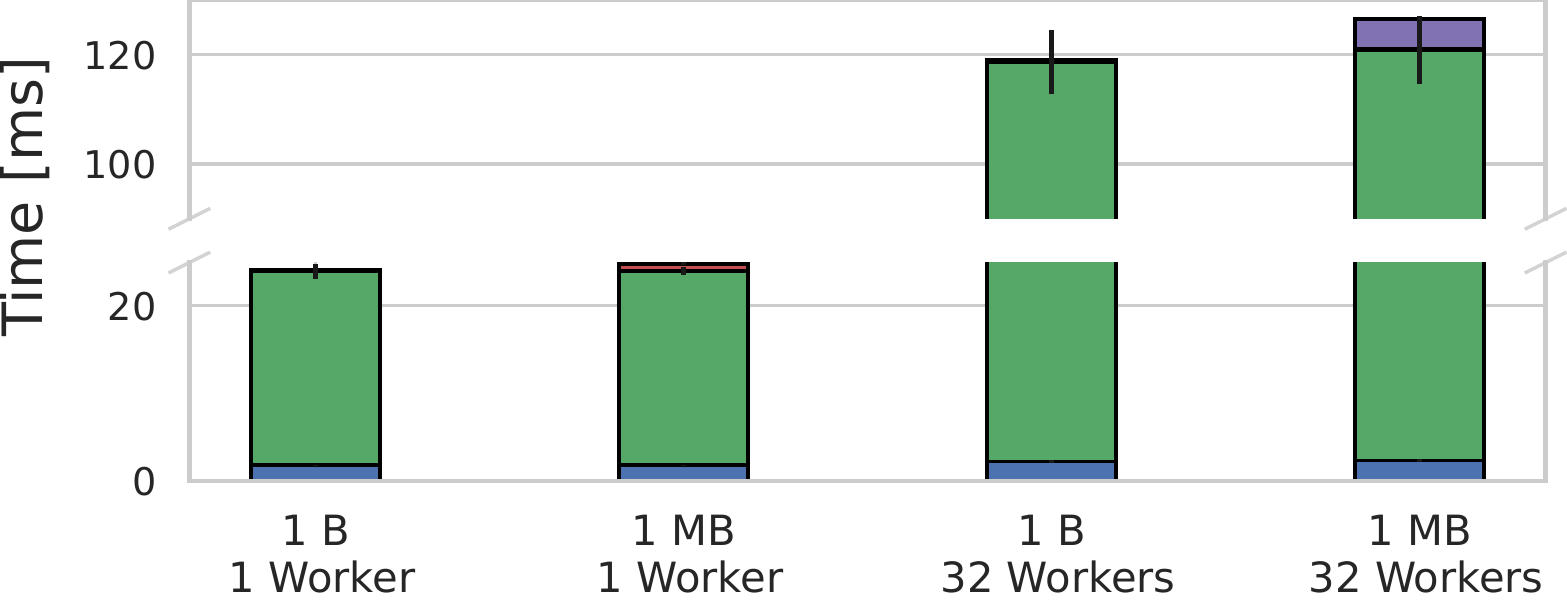}\label{fig:cold_baremetal}
    }
	}
	\hfill
	\resizebox*{0.95\width}{0.85\totalheight}{
		\subfloat[Docker container
		]{%
      \includegraphics[width=1.0\linewidth]{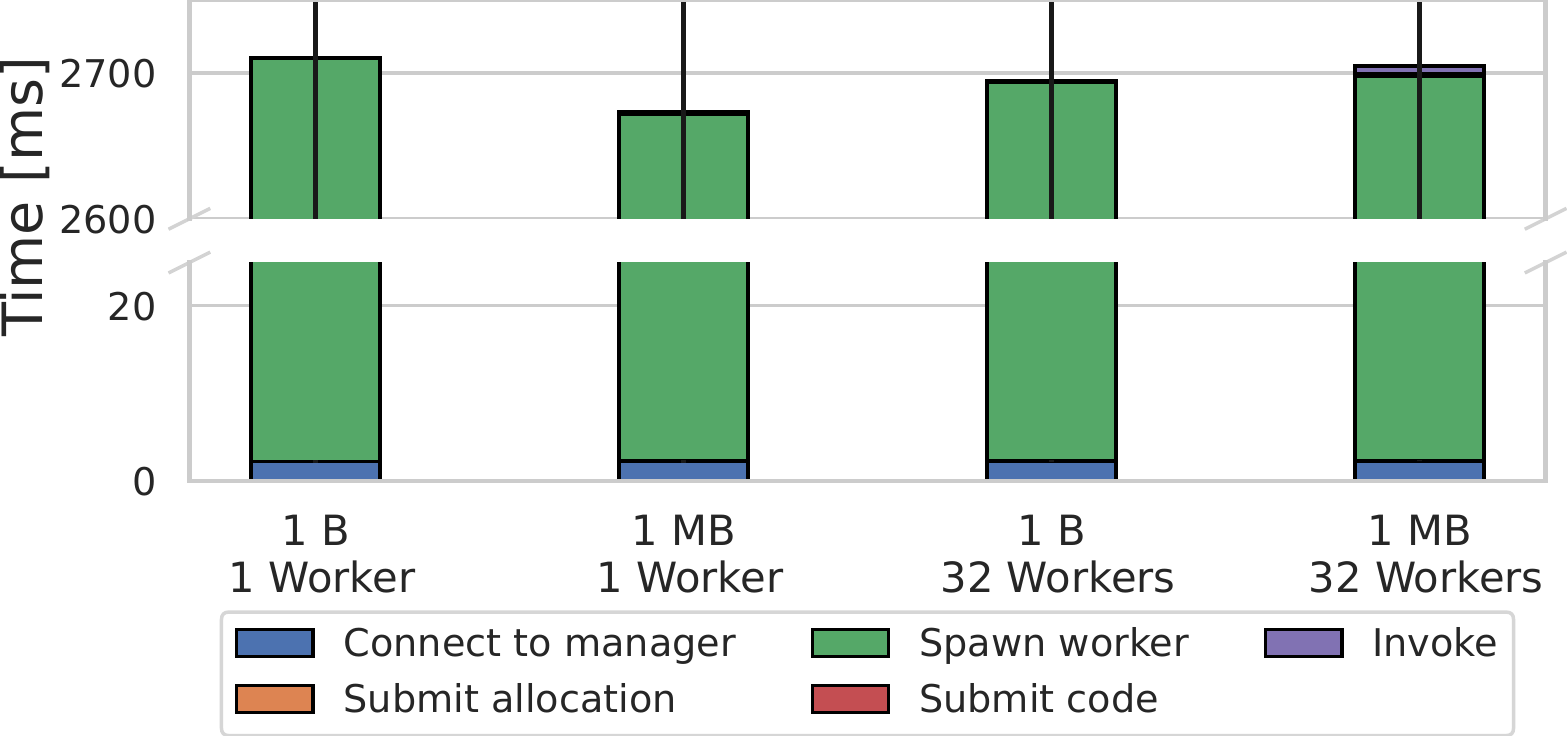}
      \label{fig:cold_docker}
		}
	}
  \caption{\textbf{Cold invocations of \toolname functions.}}
\end{figure}

Fig.~\ref{fig:cold_baremetal} and~\ref{fig:cold_docker} present the overhead of a single \emph{cold}
invocation on a bare-metal and Docker-based executor, respectively.
The data comes from 1000 invocations with a single no-op C++ function, compiled into
a shared library of size 7.88 kB.
In all tested configurations, the longest step is the creation of workers. 
All other steps: the connection establishment to the manager, submitting an allocation and code,
and code invocation, take single-digit milliseconds to accomplish.
Therefore, we can claim that \toolname{} does not introduce significant overheads in addition to sandbox initialization.

While the current version of Docker with SR-IOV shows an overhead of approximately 2.7 seconds to spawn workers,
low-latency approaches can reduce this time to as little as 125 milliseconds~\cite{246288}.
%
%
Thus, it is compatible with proposed approaches to reduce cold startup latencies, such as container warming and reinitialization.
Finally, the user's function can be deployed as a code package like in many other FaaS platforms, allowing executor managers to
keep a pool of generic and ready containers and bypass the container startup latency.
%
 
%

We therefore claim that \textbf{cold invocation overheads of \toolname do not pose an obstacle for the use in HPC}.

\subsection{Bandwidth scalability}
\label{sec:evaluation_payload}

To compare the performance of \toolname and other platforms, 
we evaluate a non-op C++ function that returns the input provided in a payload range from 1 kB to 5 MB.
Since other platforms cannot accept raw data, we generate a base64-encoded
string that approximately matches the input size.
\iftr
We compare against AWS Lambda~\cite{awsLambda}, a state-of-the-art commercial FaaS solution,
as Azure Functions~\cite{azureFunctions} and Google Cloud Functions~\cite{googleFunctions}
do not support C++ functions.
Then, we compare against open-source FaaS platforms OpenWhisk~\cite{openwhisk},
and Nightcore~\cite{nightcore}, a low-latency and open-source serverless platform.
\else
We compare with AWS Lambda, a state-of-the-art commercial FaaS solution,
as neither Azure Functions nor Google Cloud Functions support C++ functions.
Then, we compare against open-source FaaS platforms OpenWhisk
and Nightcore~\cite{nightcore}, a low-latency and open-source serverless platform.
\fi
Both open-source systems are deployed in the same RDMA-capable cluster as \toolname{}, using the same network and CPU
resources as our system.
\iftr
In Lambda, we deploy a native function implemented using the official C++ Runtime~\cite{awsCRuntime},
we expose an HTTP endpoint with no authorization, and run the experiment in
an AWS \emph{t2.micro} VM instance in the same region as the function.
We deploy on the cluster a standalone OpenWhisk using Docker with Kafka and API gateway~\cite{openwhiskDocker}.
\else
In Lambda, we deploy a native function implemented with C++ Runtime,
expose an HTTP endpoint with no authorization, and run the experiment in
an AWS \emph{t2.micro} VM instance in the same region as the function.
We deploy on our cluster a standalone OpenWhisk using Docker with Kafka and API gateway.
\fi
A C++ function in OpenWhisk is invoked as a regular application, accepting inputs not greater than 125 kB
through \emph{argc} and \emph{argv}.
Similarly, we deploy the function in a \emph{nightcore} instance in the cluster.

We present the evaluation result in Fig.~\ref{fig:evaluation_faas} (page 1).
On all payload sizes, \toolname clearly provides significantly better performance.
\toolname invocations are between 695x and 3,692x faster than AWS Lambda executions.
Stable measurements on the no-op function indicate that the difference is not caused by shared CPU resources of the cloud,
but by the low-latency network and native support for transmitting raw data.
\toolname{} is between 23x and 39x faster than Nightcore when running on the same hardware.
%
Similarly, \toolname provides a speedup between 5,904x and 22,406x when compared to OpenWhisk.

\toolname provides significant performance improvements over current FaaS platforms and \textbf{scales well with message size}.


\begin{figure}[tbh]
	\centering
  \includegraphics[width=0.98\linewidth]{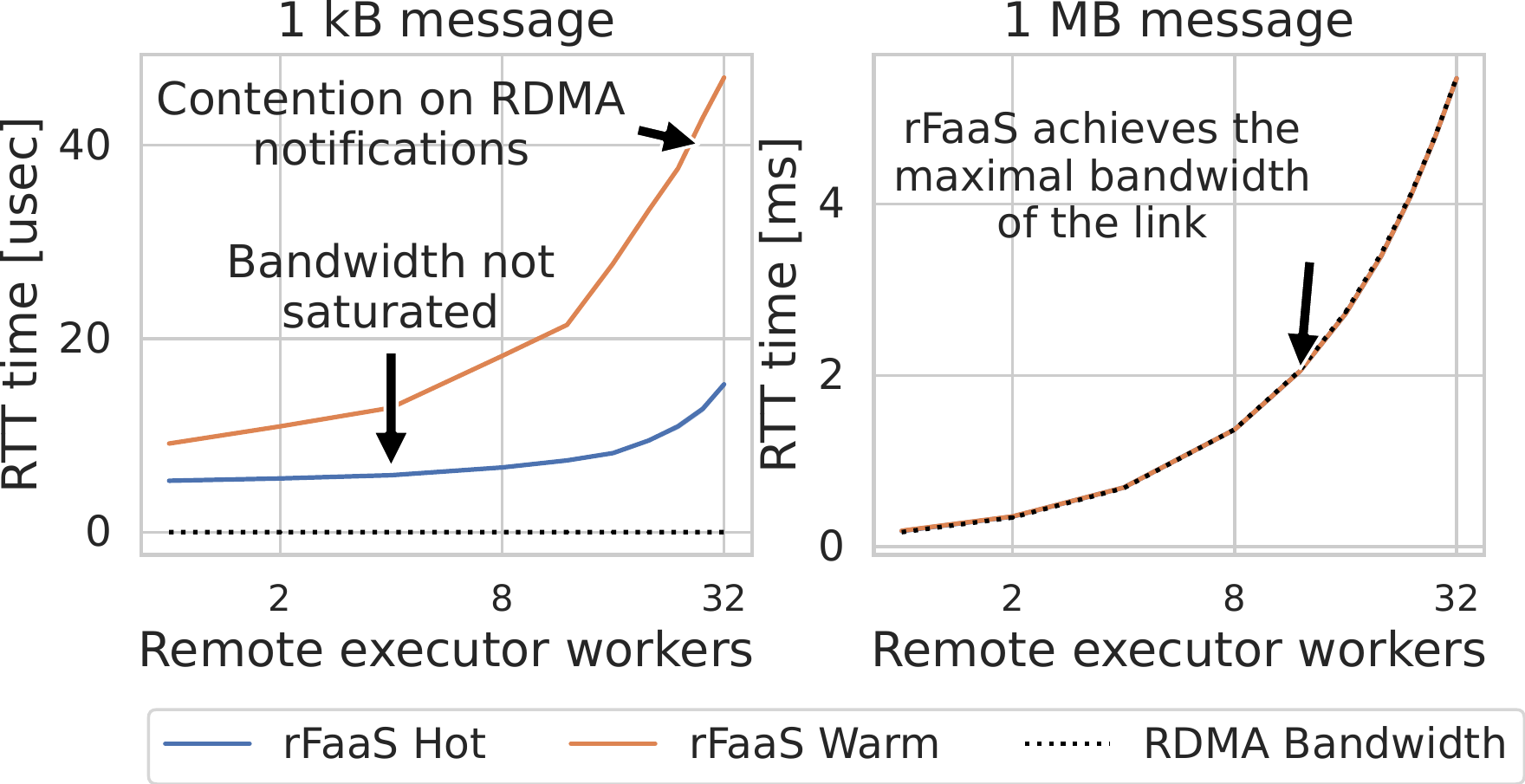}
  \caption{\textbf{\toolname invocations on parallel executors.}}
  \label{fig:parallel_combined}
\end{figure}

\subsection{Parallel scalability}
\label{sec:evaluation_parallel}

To verify that RDMA-capable functions scale efficiently to handle integration
into scalable applications, we place managers on 36-core CPUs
and evaluate parallel invocations. 
We execute the no-op function on warmed-up, bare-metal executors having allocated from 1 to 32
worker threads.

Fig.~\ref{fig:parallel_combined} presents the round-trip latencies
for invoking functions with 1 kB and 1MB payloads, respectively.
The overhead of handling many concurrent connections is insignificant on hot invocations
with a smaller payload.
While the Docker executor shows performance increases (hot) and decreases (warm)
on the 1 kB payload, the difference on 1MB payload is less than 1\%.
However, execution times increase significantly with the number of workers when sending
1 MB data, due to saturating network capacity (100 Gb/s).
This shows that \toolname scaling is limited only by the available bandwidth.

Therefore, we claim that \textbf{parallel scaling of \toolname{} executors is bounded only by network capacity}.

\begin{figure}[tb!]
	\centering
  \subfloat[Image processing: thumbnail generation: \emph{AWS Lambda} and \emph{\toolname{}}.]{
    \includegraphics[width=1.0\linewidth]{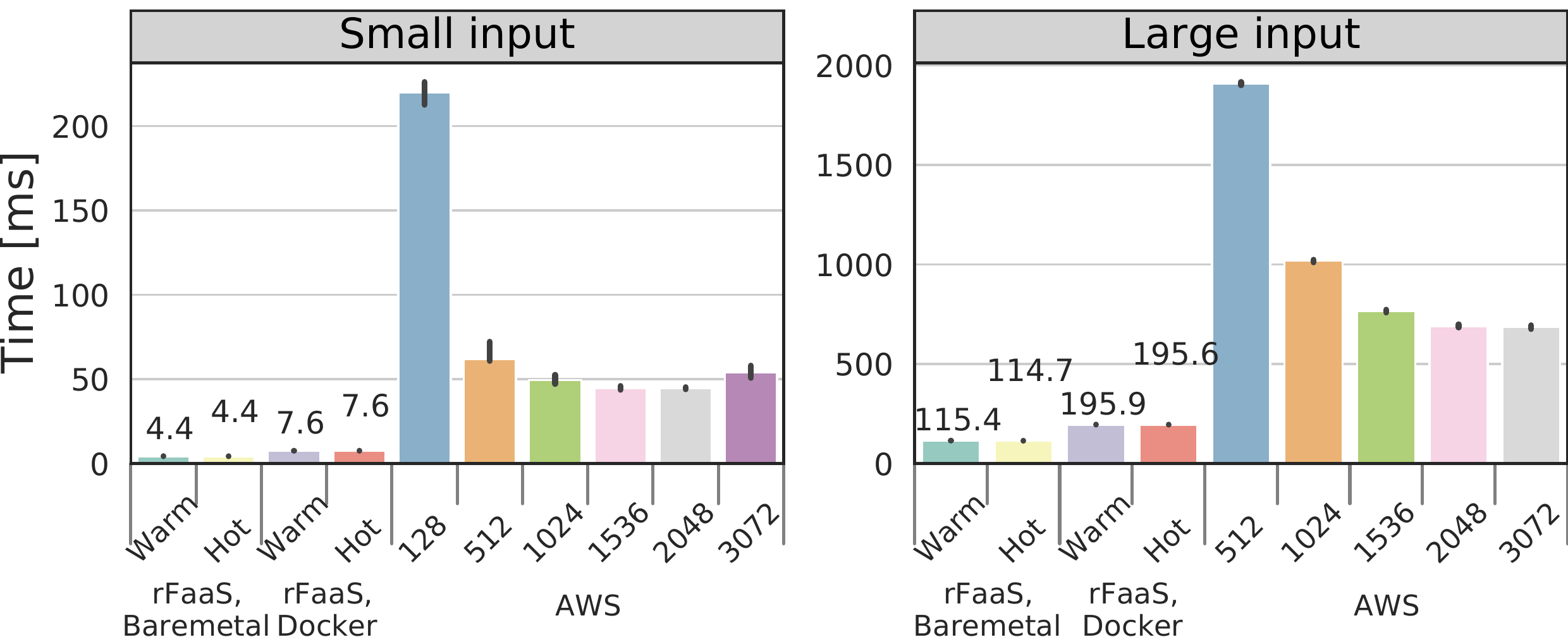}
    \label{fig:faas_thumbnailer}
  }
  \hfill
  \subfloat[Image recognition with ResNet-50 and PyTorch: \emph{AWS Lambda}, \emph{\toolname{}}.]{
    \includegraphics[width=1.0\linewidth]{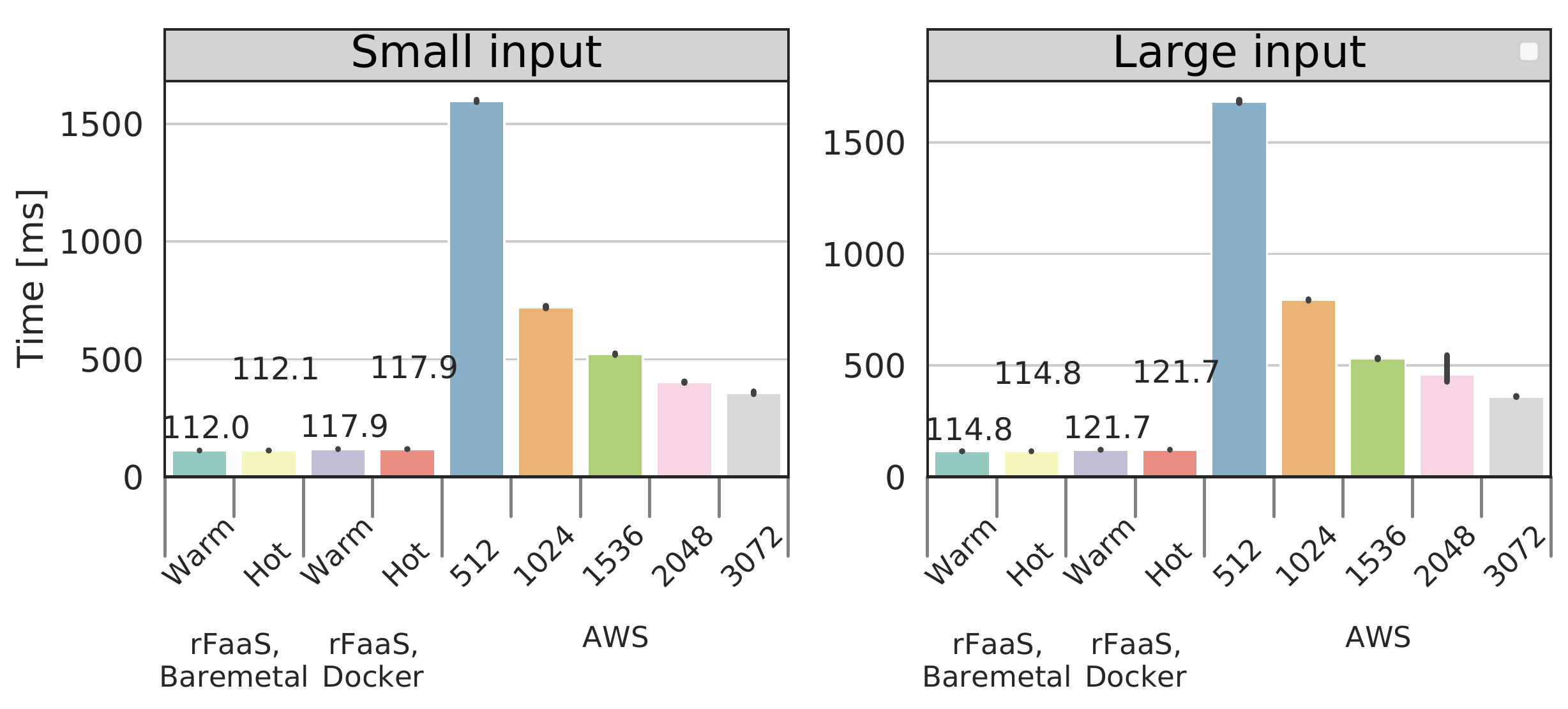}
    \label{fig:faas_recognition}
  }
  \caption{\textbf{\toolname{} on serverless functions.}}
\end{figure}

\subsection{Use-case: serverless functions}
To evaluate the effectiveness of \toolname{} in integrating serverless functions
into high-performance applications, we select real-world serverless functions from the SeBS benchmark~\cite{copik2021sebs}.
We take the \emph{thumbnailer} benchmark as an example of general-purpose image processing
and the \emph{image-recognition} benchmark performing ResNet-50 prediction as an example of integrating
deep-learning inference into applications.
We reimplement the Python benchmarks in C++ and deploy them as Docker images on \toolname{}
and AWS Lambda.
%
%
We repeat each benchmark 100 times.
\paragraph{Image processing}
We implement the thumbnail generation with OpenCV 4.5.
We evaluate functions with two images, 97 kB \emph{small} one and a 3.6 MB \emph{large} one 
(Fig.~\ref{fig:faas_thumbnailer}).
For the AWS Lambda function, we need to submit the binary image data as a base64-encoded string
in the POST request, which adds the overhead of encoding and conversions.
On the other hand, \toolname{} functions benefit from the payload format
not constrained by cloud API requirements.
\paragraph{Image registration}
We implement the benchmark with the help of PyTorch C++ API, using \emph{OpenCV} 4.5, \emph{libtorch}
1.9, and \emph{torchvision} 0.1.
We convert the Python serialized model included with \emph{SeBS} into the recommended TorchScript model format.
The model is included with the Docker image and stored in the function memory after the first invocation.
We evaluate functions with two inputs, a 53 kB \emph{small} image and a 230 kB \emph{large} one 
(Fig.~\ref{fig:faas_recognition}).
There is a growing interest in using machine-learning inference to speed up computations and
simulations~\cite{brace2021achieving,10.3389/fdata.2020.604083,Krupa_2021}.
The results demonstrate that \textbf{\toolname{} functions can efficiently implement inference
tasks in high-performance applications}.

%

\begin{figure}[t]
  \centering
  \includegraphics[width=1.0\linewidth]{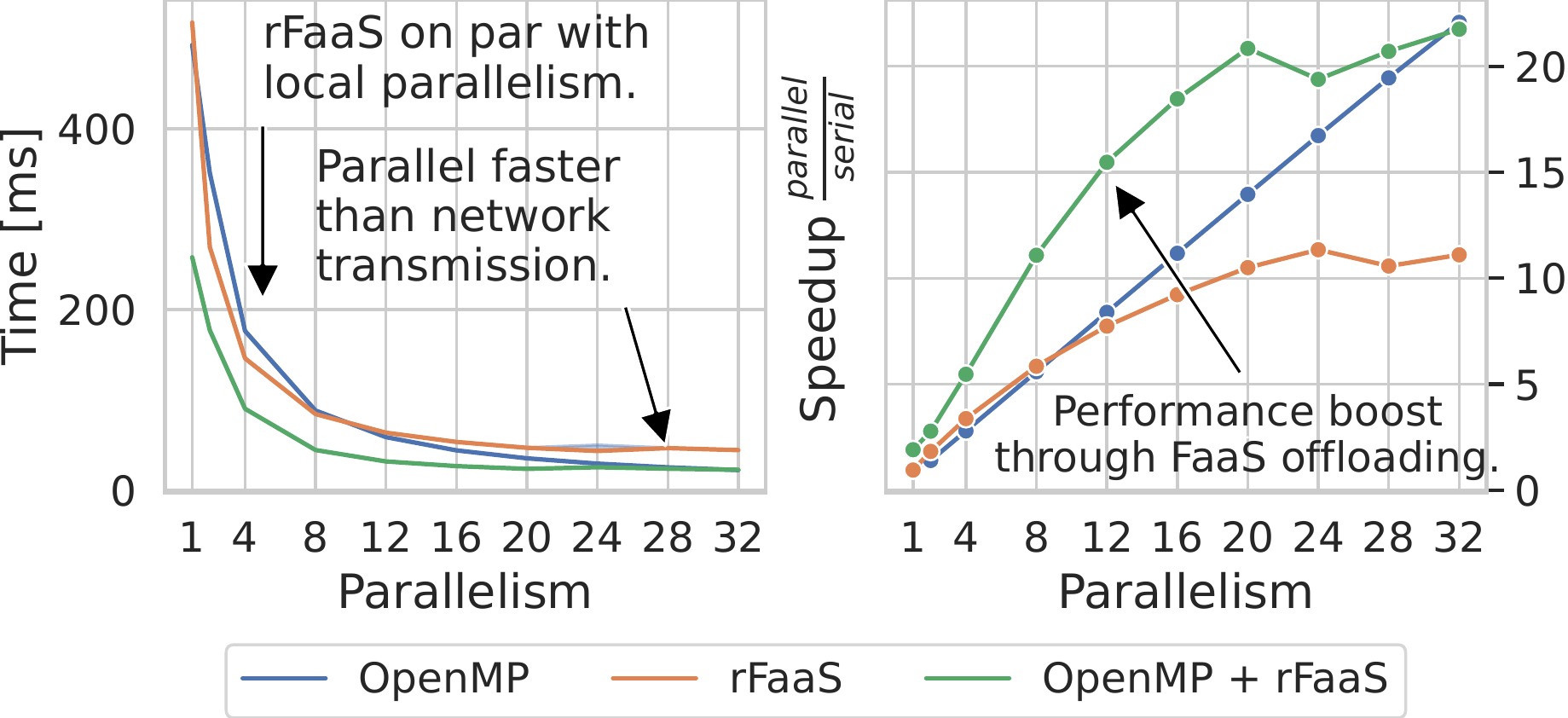}
  \caption{\textbf{Parallel serverless computing with \toolname{} and \emph{OpenMP}. Medians with non-parametric 95\% CIs.}}
  \label{fig:blackscholes}
\end{figure}

\subsection{Use-case: parallel offloading}
We want to answer the next question: can \toolname{} offload computations efficiently to remote serverless workers?
%
We study offloading of massively parallel computations with significant data movement.
We select the Black-Scholes solver~\cite{10.1080/00207160.2012.690865} from the
PARSEC suite~\cite{10.1145/1454115.1454128} parallelized with OpenMP threading.
Black-Scholes solves the same partial differential equation for different parameters,
and we dispatch independent equations to bare-metal parallel executors.
We evaluate the benchmark 
with approx. 229 MB of input and 38 MB of output
and present results in Fig.~\ref{fig:blackscholes}.
%

We show that offloading the entire work to \toolname{} scales efficiently compared to OpenMP,
as long as the workload per thread is not close to the network transmission time of approximately 20 ms.
We can further speed up the OpenMP application by offloading
half of the work to the same number of serverless functions (\emph{OpenMP + }\toolname{}).
Since other high-performance FaaS systems achieve a fraction of available bandwidth (Sec.~\ref{sec:evaluation_payload}),
their runtime will be dominated by the transmission of 229MB of data to functions.
\textbf{Thus, we can conclude that \toolname{} offers scalable parallelism bounded by network performance only.}

  
\begin{figure}[t]
 \centering
   \subfloat[Matrix-matrix multiplication.]{
    \includegraphics[width=0.47\linewidth]{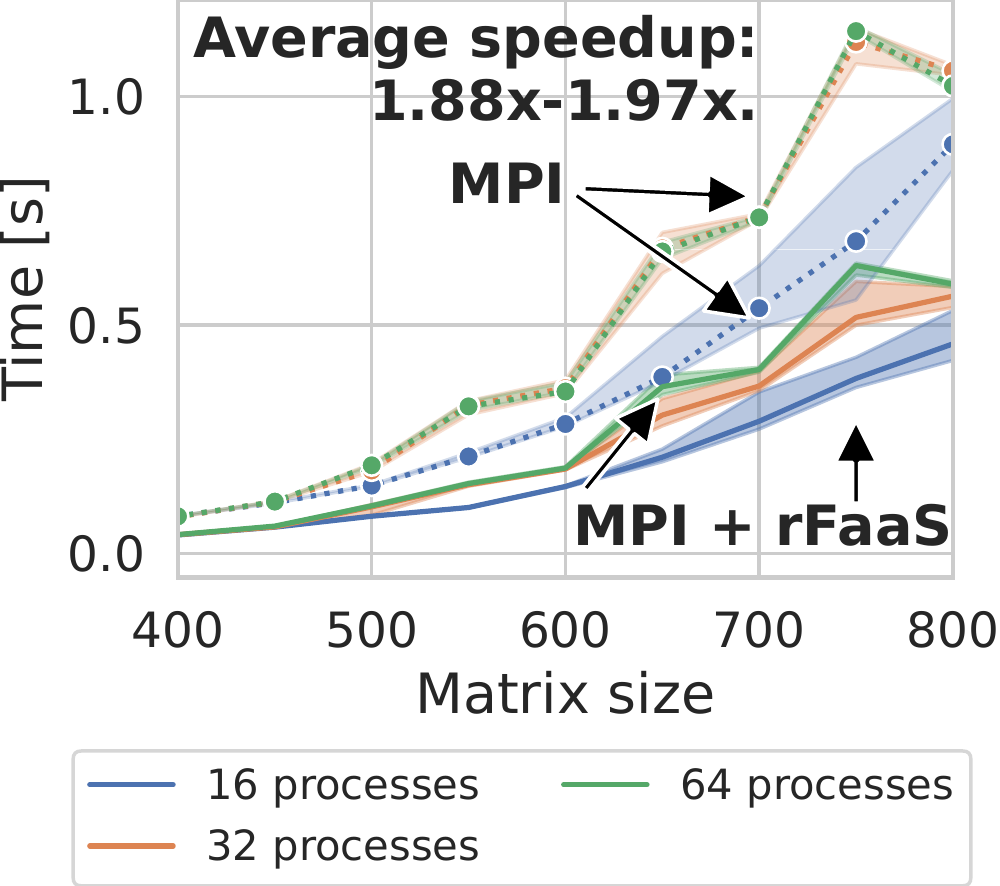}
    \label{fig:res_mmm}
  }
  \subfloat[Jacobi method, 100 iterations.]{
    \includegraphics[width=0.47\linewidth]{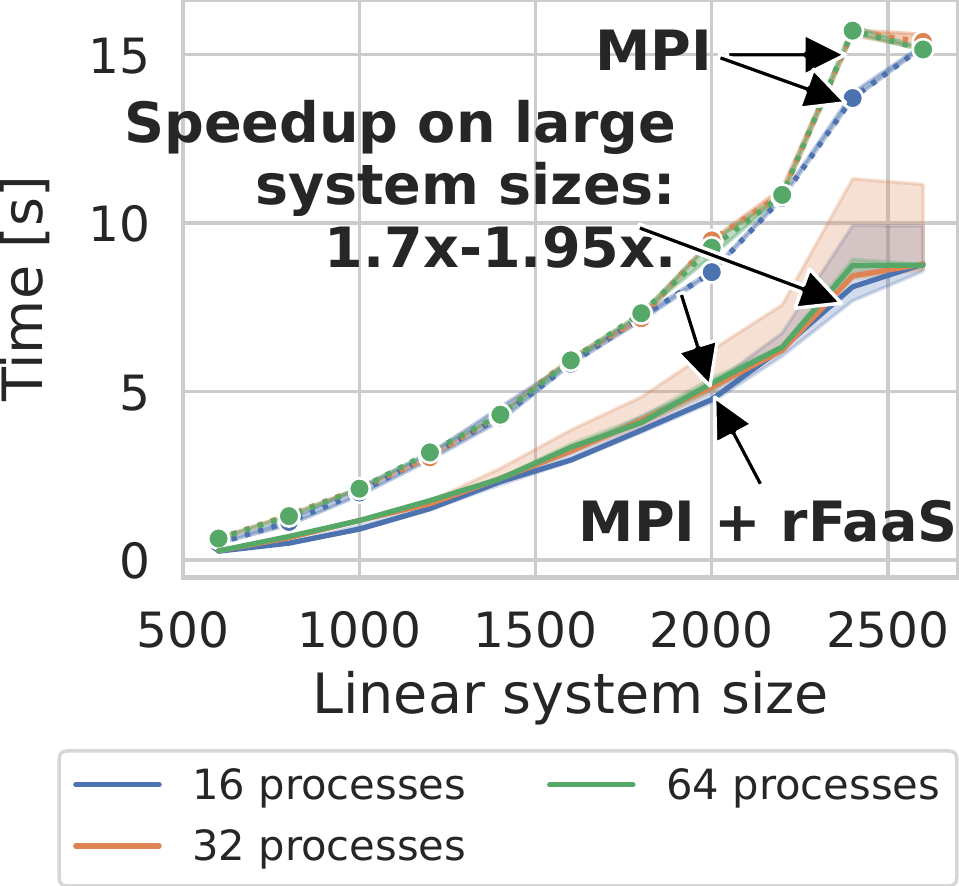}
    \label{fig:res_jacobi}
  }
  \caption{\textbf{MPI (solid) versus MPI + \toolname{} (dashed), reported medians with non-parametric 95\% CIs.}}
\end{figure}


\subsection{Use-case: HPC Applications}
%
%
%

The next question we want to answer is: how much performance can be gained by offloading
complex tasks to the cheap and spare capacity of HPC clusters?

\paragraph{Matrix-matrix multiplication}
\label{sec:eval-mat-mat}
We run an \emph{MPI} application where each rank performs a matrix-matrix multiplication,
averages it over 100 repetitions, and we measure the median kernel time across MPI ranks.
MPI ranks are distributed across two 36-core nodes, and we pin each rank to a single core.
Then, we deploy an \emph{MPI + rFaaS} application where each rank allocates a single bare-metal
\toolname{} function.
\toolname{} executors are deployed on two other 36-core nodes, and with such concentration
of MPI and \toolname{} computing resources, we show that sharing the network bandwidth does not
prevent efficient serverless acceleration.
Due to a high computation to communication ratio,
we split the workload equally, and both MPI rank and the function compute half of the result matrix.
Figure~\ref{fig:res_mmm} shows \toolname provides a speedup between
1.88x and 1.94x depending on the number of MPI processes.
%
%
%
%
%
%
Functions with a good ratio of computation to unique memory access can
be accelerated with \toolname. As long as this condition is satisfied,
\textbf{\toolname improves the performance of HPC workloads}.

\paragraph{Jacobi linear solver}
To show a serverless acceleration of a bulk synchronous type of problem,
we consider the Jacobi linear solver,
where half of each iteration is offloaded to \toolname{}.
Here, we perform a classical serverless optimization of caching resources in a warmed-up
sandbox.
Since the matrix and right-hand vector do not change between iterations, we submit
them only for the first invocation.
As long as the allocated function is not removed, we send only
an updated solution vector in subsequent iterations.

We evaluate the approach in the same setting as matrix multiplication (Section~\ref{sec:eval-mat-mat}),
with MPI ranks averaging the Jacobi method with 1000 iterations over ten repetitions, and measure
a speedup between 1.7 and 2.2 when \toolname{} acceleration is used.
Since each iteration takes just between 1 and 15 milliseconds, the
results must be returned with minimal overhead
to offer performance comparable with the main MPI process.
\textbf{the low-latency invocations in \toolname{} apply to millisecond-scale computations.}

%
%
%
%
%

\section{Related Work}
\label{sec:related_work}

\iftr

\paragraph{High-performance FaaS}
FuncX~\cite{funcx} is a federated and distributed FaaS platform designed to bring serverless
function abstraction to scientific computing.
Nonetheless, FuncX does not take advantage of high-speed networks and implements a hierarchical
and centralized design with long invocation paths between clients and remote workers.
As a result, even warm invocations take at least 90ms.
Nightcore~\cite{nightcore} is a high-performance FaaS runtime that optimizes internal function calls
--- invocations made by a running function that can be satisfied locally, without inter-node communication.

Cloudburst~\cite{10.14778/3407790.3407836} brings stateful computations and consistency into
serverless workflows with auto-scalable key-value storage.
SAND~\cite{10.5555/3277355.3277444} is a serverless platform optimized for workflows of serverless
functions through a grouping of functions and dedicated message buses for subsequent invocations.
In contrast, \toolname{} exploits co-location via explicit parallelism of executor
allocation and optimizes invocation latencies through RDMA communication.
Archipelago~\cite{archipelago} and Wukong~\cite{wukong} allow users to submit jobs that are
represented as a directed acyclic graph (DAG) of functions. 
They perform latency-aware scheduling of a submitted DAG: Wukong uses a decentralized and dynamic
scheduling built on top of AWS Lambda, while Archipelago focuses on resource partitioning for
decentralized schedulers and optimizing the control plane. 
%
In \toolname{} both allocation and invocation are decentralized and optimized with
a direct client-worker connection.

The improvements and optimization strategies,
such as the application-layer solutions in Wukong,
sandbox warming up in Archipelago,
container provisioning system SOCK~\cite{216031},
and the fast startup sandbox Catalyzer~\cite{10.1145/3373376.3378512},
are orthogonal to \toolname{} and can be implemented in our platform as well.

SmartNICs have been shown to provide fast dispatching and orchestration for FaaS platforms~\cite{10.1145/3472883.3486982}.
Choi et al. presented low-latency invocations on a SmartNIC runtime~\cite{9355790}.
However, supported functions are limited by restricted C-like implementation language and low-performance RISC cores.

\paragraph{Remote Invocations}
Remote Procedure Calls (RPC)~\cite{nelson1981remote} and Active Messages~\cite{eicken} provide the ability to invoke a
procedure remotely on another machine.
Active Networks include capsules with user code that can be executed on selected routers~\cite{10.1145/319151.319156}.
In comparison, \toolname{} provides the elasticity of executing on dynamically allocated resources with the pay-as-you-go billing instead
of requiring provisioned resources.
We enable multi-tenant computations on a single server by providing isolation.
Since \toolname{} do not send code with invocation, we provide a protection boundary between caller and callee
needed to access private cloud resources.

\paragraph{Boosting utilization}
Many approaches attempt to boost the utilization of cloud and cluster resources~\cite{10.1145/2741948.2741964,186175,hindman2011mesos,10.1145/2644865.2541941}.
%
These are focused on reclaiming idle resources and co-locating offline batch jobs with online
and latency-sensitive cloud services, and these approaches are orthogonal to \toolname{}
as they cannot target ephemeral resources efficiently.

Zhang et al.~\cite{10.1145/3477132.3483580} implement an OpenWhisk load balancer optimized for harvested idle resources.
This method targets centralized FaaS platforms and does not provide high-performance invocations.

\else
FuncX~\cite{funcx} is a federated platform that brings function abstraction to scientific computing.
Nonetheless, FuncX does not take advantage of high-speed networks and implements a hierarchical
and centralized design with long invocation paths between clients and remote workers.
As a result, even warm invocations take at least 90ms.
Nightcore~\cite{nightcore} is a high-performance FaaS runtime designed for microservices,
with optimized internal function calls --- invocations that can be satisfied locally, without inter-node communication.
%
SAND~\cite{10.5555/3277355.3277444} optimizes workflows of serverless
functions with grouping of functions and dedicated message buses.
In contrast, \toolname{} exploits co-location via explicit parallelism of executor
allocation and optimizes invocation latencies through RDMA communication.
Archipelago~\cite{archipelago} and Wukong~\cite{wukong} perform latency-aware scheduling
of directed acyclic graph (DAG) of functions. 
Wukong uses a decentralized and dynamic scheduling built on top of AWS Lambda, while Archipelago
focuses on resource partitioning for decentralized schedulers and optimizing the control plane.
%
In \toolname{} both allocation and invocation are decentralized and optimized with
a direct client-worker connection.
SmartNICs have been shown to provide fast dispatching and orchestration in FaaS~\cite{10.1145/3472883.3486982,9355790}.
However, functions are limited by restricted implementation language and low-performance RISC cores.
Other improvements and optimization strategies, such as warming, provisioning and fast startup
solutions for sandboxes~\cite{216031,10.1145/3373376.3378512},
are orthogonal to \toolname{} and can be implemented in our platform as well.
RDMA has been used in the serverless context for resource disaggregation~\cite{guo2022resourcecentric}
and heterogeneous systems~\cite{10.1145/3503222.3507732}; \toolname{} optimizes FaaS architecture and is tailored for HPC computing.

\textbf{Remote Invocations}
Remote Procedure Calls (RPC)~\cite{nelson1981remote} and Active Messages~\cite{eicken}
invoke a procedure remotely on another machine.
Active Networks include capsules with user code that can be executed on selected routers~\cite{10.1145/319151.319156}.
In comparison, \toolname{} provides the elasticity of executing on dynamically allocated resources with the pay-as-you-go billing instead
of requiring provisioned resources.
We enable multi-tenant computations on a single server by providing isolation.
Since \toolname{} does not send code with invocation, we provide a protection boundary between
caller and callee needed to access private resources, e.g., in ML inference serving.

\fi

\section{Discussion}

In this paper, we introduce RDMA abstractions into FaaS to facilitate
the integration of functions into high-performance and latency-sensitive applications.
\toolname{} can positively impact other aspects of serverless systems,
and we now discuss how our protocols combine with other emerging solutions in FaaS.

\textbf{Which workloads will benefit from \toolname{}?}
%
%
High-performance and parallel applications need scalable invocations of remote workers (Sec.~\ref{sec:evaluation_parallel})
and high network bandwidth to support simultaneous invocations by parallel processes on the same node (Sec.~\ref{sec:evaluation_payload}).
Furthermore, data-intensive workloads will benefit from RDMA-accelerated FaaS computing
since other platforms cannot achieve high throughput on networks that support the transmission of
gigabytes of data per second.
Examples include, but are not limited to, machine-learning inference, data analytics,
GPU-accelerated functions with short computation time, and task-based applications with
no memory sharing between tasks~\cite{copik2020workstealing}.

On the other hand, HPC applications that will likely not benefit from \toolname{} offloading include
memory-bound operations with a low ratio of computation to accessed data.
Furthermore, applications that already achieve high resource utilization have little motivation
to look into serverless in the first place, e.g., applications with static parallelism and homogenous
resource requirements.

%
%
%
%

%
%

\textbf{Can \toolname{} improve serverless workflows?}
In workflows, functions are composed to build serverless applications, using a coordination
service to orchestrate invocations and data propagation~\cite{10.1145/3485510}.
While SmartNICs offer fast orchestration~\cite{10.1145/3472883.3486982}, they are
limited by the cost and availability of dedicated NICs.
Instead, implementing orchestrator with \toolname{} executors achieves two performance goals:
single-digit microsecond latency overhead of invocations and efficient data movement.


\textbf{Can \toolname{} support the diverse world of HPC systems?}
Thanks to its modular design, \toolname{} supports extensions into new environments and hardware.
\toolname{} functionality and RDMA abstractions are orthogonal to the device interface,
and network management with \code{ibverbs} can be extended with new network drivers
and software virtualization for RDMA~\cite{225978}.
\toolname{} supports native integration into C/C++ applications (Sec.~\ref{sec:integration}),
but the language choice is independent of the platform itself.
Functions and integration can be implemented through ABI compatibility
and foreign-function interfaces, supporting languages such as Fortran and Python.
\toolname{} functions can be served in containers other than Docker,
e.g., in HPC container Singularity, as long as they provide access to the RDMA NIC.

\section{Conclusions}
Fine-grained and granular computing need systems designed to handle microsecond-scale workloads~\cite{lopez2021serverless,10.1145/3317550.3321447},
but FaaS platforms still operate at the millisecond latency.
\toolname{} attempts to solve this problem at three levels:
a novel direct and decentralized scheduling to reduce serverless critical path,
incorporation of high-speed networks to achieve microsecond-latency,
and inclusion of remote memory access to remove overheads of the OS control plane.
With RDMA-capable functions, we demonstrate hot invocations with less than
one microsecond of overhead and efficient parallel scalability,
providing serverless programmability in high-performance systems and applications,
and paving the way for future low-latency and fine-grained computing.

\section*{Acknowledgment}

  \begin{wrapfigure}{r}{.13\linewidth} 
    \includegraphics[width=.13\columnwidth]{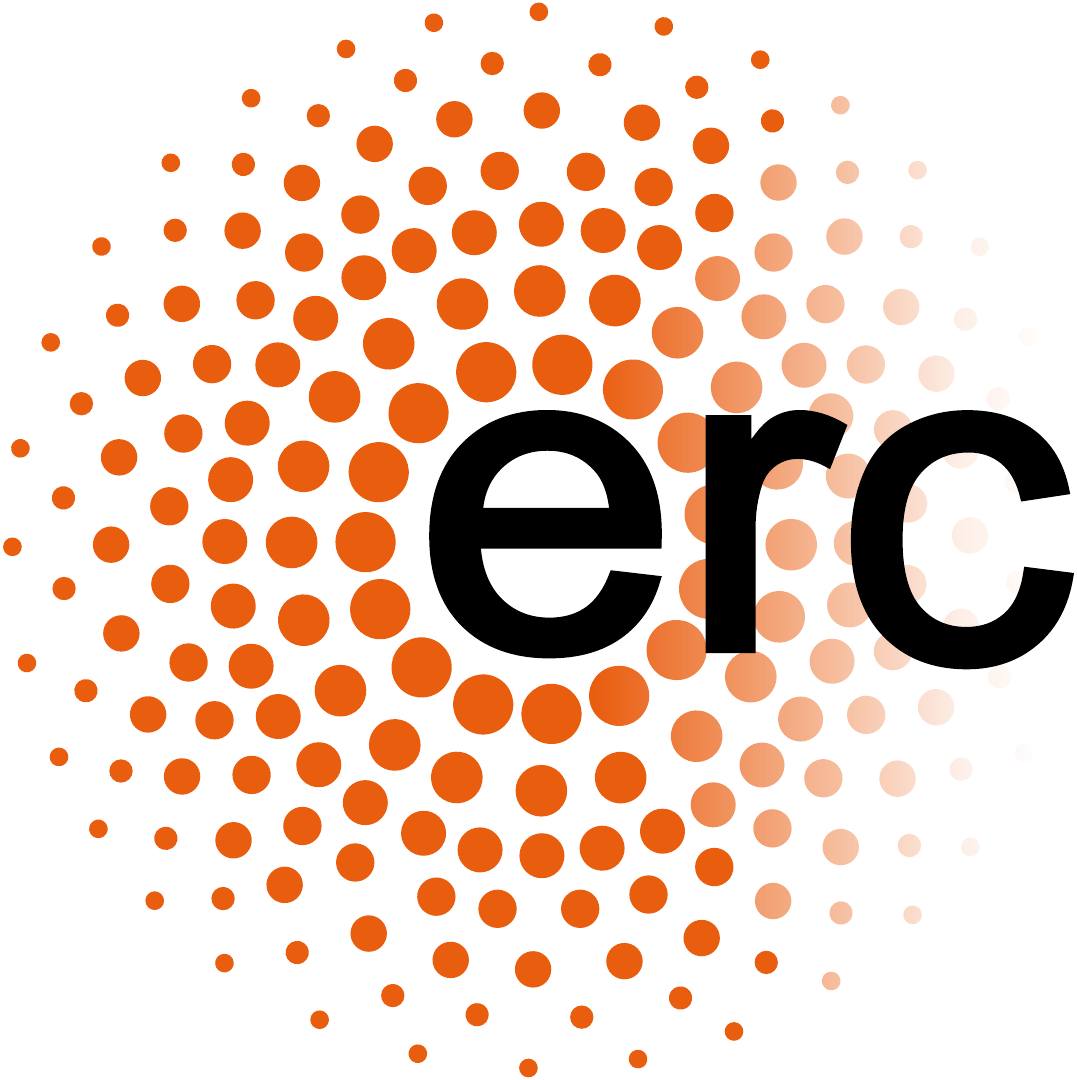}
  \end{wrapfigure}

  This project has received funding from the European Research Council (ERC) under the
  European Union’s Horizon 2020 programme (grant agreement EPIGRAM-HS, No. 801039,
  and grant agreement RED-SEA, No. 955776),
  and from the Schweizerische Nationalfonds zur Förderung der
  wissenschaftlichen Forschung (SNF, Swiss National Science Foundation) through Project 170415.
  We would also like to thank the Swiss National Supercomputing Centre (CSCS) for providing us with
  access to their supercomputing machines Daint and Ault.
  We also thank anonymous reviewers for helping us improve the manuscript.

  \bibliographystyle{IEEEtran}
  \bibliography{serverless,cloud,paper,rdma}

\end{document}